\documentclass[12pt]{article}

\usepackage{jheppub}
\makeatletter
\def\@fpheader{\relax}
\makeatother

\usepackage{caption}
\usepackage{subcaption}
\usepackage{amsmath,amssymb,amsfonts}
\usepackage{graphicx}
\usepackage{makeidx}
\usepackage{multicol}
\usepackage{multirow}
\usepackage{float}
\usepackage[table]{xcolor}
\usepackage{hyperref}
\usepackage{makecell}

\newcommand\be{\begin{equation}}
\newcommand\ee{\end{equation}}
\newcommand\bea{\begin{eqnarray}}
\newcommand\eea{\end{eqnarray}}
\newcommand\ba{\begin{array}}
\newcommand\ea{\end{array}}
\newcommand\cI{\mathcal{I}}
\newcommand\cS{\mathcal{S}}
\newcommand\cR{\mathcal{R}}
\newcommand\cD{\mathcal{D}}

\newcommand\comment[1]{}

\begin{document} 

\title{Time-dependent and time-independent SIR models applied to the COVID-19 outbreak in Argentina, Brazil, Colombia, Mexico
and South Africa }


\author[a,b]{\!\!, Nana Geraldine Cabo Bizet}
\author[c,b]{\!\!, Dami\'an Kaloni Mayorga Pe\~na}

\affiliation[\,a]{Departamento de Física, DCI, Universidad de Guanajuato,
Loma del Bosque No. 103 Col. Lomas del Campestre C.P 37150 Leon, Guanajuato, Mexico}
\affiliation[\,b]{Data Laboratory, Universidad de Guanajuato,
Loma del Bosque No. 103 Col. Lomas del Campestre C.P 37150 Leon, Guanajuato, Mexico}
\affiliation[\,c]{Mandelstam Institute for Theoretical Physics, School of Physics, NITheP, and CoE-MaSS, \\ University of the Witwatersrand, Johannesburg, WITS 2050, South Africa}

\emailAdd{nana@fisica.ugto.mx}
\emailAdd{damian.mayorgapena@wits.ac.za}

\date{\today}

\abstract{

We consider  the SIR epidemiological model applied to the evolution of COVID-19 with two approaches. In the first place we fit a global SIR model, with time delay, and constant parameters throughout the outbreak, including the contagion rate. The contention measures are reflected on an effective reduced susceptible population $N_{eff}$. In the second approach we consider a time-dependent contagion rate that reflects the contention measures either through a step by step fitting process or by following an exponential decay. In this last model the population is considered the one of the country $N$. In the linear region of the differential equations, when the total population $N$ is large the predictions are independent of $N$.  We apply these methodologies to study the spread of the pandemic in 
Argentina, Brazil, Colombia, Mexico, and South Africa for which the infection peaks
are yet to be reached. In all of these cases we provide estimates for the reproduction and recovery rates.
The scenario for a time varying contagion rate is optimistic, considering that reasonable 
measures are taken such that the reproduction factor $R_0$  decreases exponentially. 
The measured values for the recovery rate $\gamma$ ​​are reported finding a universality of this parameter over various countries.  We discuss the correspondence between the global SIR with effective population $N_{eff}$
and the evolution of the time local SIR. }



\maketitle


\section{Introduction}

In this year we have been confronting  a very challenging situation in the world with the expansion of
COVID-19 virus. Its contention remains a challenge to most of the countries and as of now, we have to rely on efficient testing, social distancing measures and other sorts of non-pharmacological intervention (NPI). Similarly, to the extent of each country's capacity and the international aid the hospital bed and ICU capacity have to be increased. 
Modeling the evolution of the pandemic becomes relevant for all of the issues mentioned above. Testing and following of the infections provides the primary data source to model the evolution of the pandemic. Prompt reporting and representative testing naturally reflect on the quality of the forecast models and the accuracy of the predictions. 
Similarly, both models and data permit us to evaluate the efficiency of the NPIs and their role in changing the course of the outbreak. 
Finally the modeling of the outbreak can provide guidelines in terms of readiness of the healthcare systems in such a way that once the infection peak is reached, all possible measures to ensure efficient attention as well as risk mitigation procedures are adequately put in place.


The SIR model is among the classics of epidemiological modeling and owes its name to the split of the population into three categories: susceptible (S), infected (I) and recovered (R). Despite of its simplicity, it has proven its versatility in the modeling of a variety of infectious outbreaks, among these, the novel coronavirus \cite{1,2,3,4,5,6,7,8,9,10,11,12,13,14,15,16,17,18,19,20,21,22,23,24,25,26}. 
The SIR model is described in terms of a system of coupled differential equations for the evolution of $S$, $I$ and $R$, and the solutions will depend on the initial conditions as well as on the contagion rate (denoted by $\beta$) as well as the recovery rate (denoted by $\gamma$). In this work we consider the SIR model and variations of it in order to describe the short term evolution of the pandemic in various latin-american countries as well as South Africa. Part of our working strategy is to follow a self contained approach, in such a way that the parameters involved in the evolution of the outbreak in each case study are automatically derived from a best fit on the data. This means that for each country we obtain estimates for $\beta$ and $\gamma$ as well as the basic reproduction number $R_0$. This is advantageous in comparison to other studies where the evolution is described upon the use of data from countries that have already navigated through the first peak, in the sense that the case by case parameters account for particular features of the population under consideration, such as mean population age, frequency of comorbidities and healthcare capacity among others. In this sense this work accounts for a self contained study based on the country by country data provided by national authorities to the World Health Organization.  However we obtain a certain universality in the parameters studied, for example the recovery rate $\gamma$ is always around 0.06 as noticed in \cite{26} from fitting the tale of the active cases curve in the case of China.


Our strategy is twofold: In the first case we consider a SIR model with time independent parameters $\beta$, $\gamma$ as well as an effective population $N_{eff}$ and a time shift $\delta t$ in the evolution. These parameters are found upon marginalization of the corresponding models to the data. Even though it is expected for the parameter $\beta$ to decrease due to the diverse NPIs put in place, this effect is reflected on a smaller value for $\beta$ as it averages over the corresponding time evolution, and more importantly on a small value of $N_{eff}$ as the NPIs tend to reduce the exposure of the overall population. This can
occur due to the characteristics of the initial spreading and also due to the
contention measures in the society.  Another issue encountered when modeling the data has to do with the fact that one usually finds an overall delay in the evolution of the curves, once this suitable time delay is included we obtain a proper fit\footnote{Models with time delay for evolution of epidemics have been discussed
in refs. \cite{7,8,9} among others.}. The cause of this delay can be explained by the fact that
the time of incubation is finite, and therefore the time variation of the different
populations should be related with the values of the magnitudes at a previous
time. There are previous studies which consider that the total susceptible
population  is a quantity which is smaller than the country (or city) population \cite{1}.


In the second case we allow for an explicit time dependent contagion rate $\beta(t)$. 
Time dependent contagion rate has been considered in previous COVID-19 studies \cite{7}. In particular in \cite{24} step, exponential and linear dependences were explored. The motivation to explore an exponential decay in  \cite{24}  is that the reduction of the contagion rate is obtained after lockdown
due to the fact that contacts are reduced to closest family members, but the disease has a life time $\approx 1/\gamma$.  However this can change if the country relaxes the contention measures.  The model was successfully applied to describe the evolution of the virus in Cuba, estimating correctly the peak of active cases.
We consider for this work  the mentioned exponential
dependence. Also in \cite{25} this dependence has been applied successfully to describe the outbreak in Italy.  Also the step dependence of $\beta$ is very accurate to study
the evolution of the epidemia, and we employ it, but it doesn't allow to predict stricter control measures,
as one can not estimate what will be the next down jump of this magnitude. In the
case when the control measures have already ensured that the reproduction number
$R_0<1$ one can extrapolate with the achieved constant contagion rate $\beta_{today}$,
to bound a worst case scenario. As the studied countries are not yet in this phase,
we take into account the functional exponential dependence.
We fit the best values of $\gamma, \beta(t)$ to a time dependent SIR considering that
the total population of susceptible is the country population $N$. This
consideration makes the SIR differential equations linear, and the
populations dependence is then locally in time an exponential.

\begin{table}[h]
\centering
\renewcommand{\arraystretch}{1.4}
{\scriptsize
\begin{tabular}{|c|c|c|c|c|}\cline{2-5}
\multicolumn{1}{c|}{} &\makecell{ Active Infection \\ Peak Day}  & \makecell{ Maximum Active \\ Infections}   &\makecell{Confirmed \\ Cases} & \makecell{Deaths} \\ \hline
 \multicolumn{5}{|c|}{ARGENTINA} \\ \hline\hline			
 $TI$ &\makecell{01.07.2020\\ (23.06.2020 - 25.07.2020)} & \makecell{14.550 \\(8.136 - 51.814)} & \makecell{108.182\\(72.568 - 361.347)} & \makecell{13.620\\(4.555 - 41.632} \\ \hline
$R_0^{(1)}$ &\makecell{ 09.06.2020 \\ (6.06.20 - 21.07.20)} & \makecell{14.205 \\ (13.010 - 29.227)}& \makecell{34.930\\ (23.489-100.446)}& \makecell{3.729\\(2.508-10.714)} \\ \hline
$R_0^{(2)}$ & \makecell{10.06.20 \\ (06.06.20 - 30.07.20)} &  \makecell{14.068 \\ (13.954 - 34.710)} & \makecell{36.707\\ (26.584 - 125.138)} &  \makecell{3.918\\ (2.838-13.335)}
\\ \hline\hline
 \multicolumn{5}{|c|}{BRAZIL} \\ \hline \hline
  $TI$ &\makecell{12.06.2020\\(03.06.2020 - 20.06.2020)} & \makecell{70.944\\(61.473 - 87.987)} & \makecell{$1.818\cdot 10^6$\\ $(1.640\,\,-\,\,2.164)\cdot 10^6$} & \makecell{224.621\\(193.658 - 272.378)} \\ \hline
$R_0^{(1)}$ &\makecell{ 14.06.20 \\ (06.06.20 - 11.07.20)} & \makecell{354.044\\(296.443 - 769.079)}& \makecell{$1.598\cdot10^6$\\ $(0.966\,\,-\,\,4.275)\cdot 10^6$}& \makecell{ 207.189\\(125.296-554.159)} \\ \hline
$R_0^{(2)}$ & \makecell{22.06.20 \\ (06.06.20 - 24.07.20)} &  \makecell{370.501\\(305.216 - $1.313\cdot 10^6$)} & \makecell{$1.94\dot10^6$ \\$(1.11\,\,-\,\,7.61)\cdot 10^6$} &  \makecell{252.096\\ (143.974-986.868)}
\\ \hline \hline
 \multicolumn{5}{|c|}{COLOMBIA} \\ \hline\hline
  $TI$ &\makecell{23.06.2020\\(15.06.2020 - 03.07.2020)} & \makecell{18.484\\(9.857 - 32.509)} & \makecell{121.851\\(75.299 - 204.447)} & \makecell{11.897\\(6.301 - 19.951)} \\ \hline			
$R_0^{(1)}$ &\makecell{20.06.20\\ (15.06.20 - 19.07.20)} & \makecell{22.006\\ (18.167 - 51.160)} & \makecell{78.839\\ (46.397 - 238.397)} & \makecell{8.465\\(4.982-25.595)} \\ \hline
$R_0^{(2)}$ & \makecell{20.06.20\\ (15.06.20 - 20.08.20)} &  \makecell{23.235\\ (22.455,113.162)} & \makecell{74.833\\(92.692 - 637.010)} &  \makecell{9.950 \\(8.034-67.533)}
\\ \hline\hline
 \multicolumn{5}{|c|}{MEXICO} \\ \hline\hline
  $TI$ &\makecell{10.06.2020\\(03.06.2020 - 22.06.2020)} & \makecell{14.349\\(8.631 - 22.552)} & \makecell{213.089\\(153.335 - 280.227)} & \makecell{30.260\\(19.311 - 42.035)} \\ \hline 			
$R_0^{(1)}$ &\makecell{22.06.20\\ (7.06.20 - 10.07.20)} & \makecell{20.749 \\(15.927 - 40.398)} & \makecell{267.474\\(182.762 - 513.906)} & \makecell{36.786\\(25.136-70.678)} \\ \hline
$R_0^{(2)}$ & \makecell{22.06.20 \\ (07.06.20 - 11.07.20)} &  \makecell{20.832 \\ (15.625 - 34.305)} & \makecell{249.115\\(182.008 - 437.077)} &  \makecell{29.255 \\(21.914- 50.499)} 
\\ \hline\hline
 \multicolumn{5}{|c|}{SOUTH AFRICA} \\ \hline\hline
   $TI$ &\makecell{15.07.2020\\(02.07.2020 - 25.07.2020)} & \makecell{14.349\\(8.631 - 22.552)} & \makecell{460.977\\(231.892 - 833.388)} & \makecell{22.949\\(5.416 - 45.242)} \\ \hline 			
$R_0^{(1)}$ &\makecell{19.06.20\\ (07.06.20 - 14.08.20)} & \makecell{21.248 \\(17.370 - 302.734)} & \makecell{141.740\\(61.777 - $2.4628\cdot 10^6$)} & \makecell{5.523
\\(2.407-95.961)} \\ \hline
$R_0^{(2)}$ & \makecell{28.06.20\\ (6.06.20 - 11.08.20)} &  \makecell{27.127\\(21.827 - 281.918)} & \makecell{234.355\\(89.251 - $2.2831\cdot10^6$)} &  \makecell{9.132\\
(3.478-88.961)} 
\\ \hline\hline
\end{tabular}
}
\caption{\textit{Day of the peak, maximum number of active cases, confirmed cases and accumulated deaths for various countries and different methodologies $TI$ are the values for the time independent SIR model, $R_0^{(1)}$ and $R_0^{(2)}$ are results for the two different time dependent fits. 
The data are obtained from the time interval 16.03.20 to 5.06.20.\label{picos}}} 
\end{table}

Both of these approaches allow us to make an estimate of when the first peak will occur for each country,  as well as the  detected active cases at the peak, the total number of detected infected and the number of deceased patients during the course of the first peak. This quantities are reported for the different models in Table \ref{picos} were we include the prediction for each methodology and the corresponding confidence intervals.

Our paper is organized as follows: In section \ref{sec:sir} we describe the  general characteristics of the SIR model. In subsection \ref{s21}  the relation of the SIR model with the logistic curve is discussed. In section \ref{s22} we discuss the reproduction number for the SIR model with constant parameters and also with variable parameters. Section \ref{s3} is devoted to the two different forecast methodologies. Subsection \ref{s31} discusses the SIR model with time independent parameters, effective population and time delay, subsection \ref{s32} describes the SIR model with time dependent $\beta(t)$ calculated by steps, while subsection \ref{s33} describes the SIR model extrapolated with an exponential
 $\beta(t)$. In the last two cases the population is the complete country population. In Section \ref{s4} we discuss the two different approaches and present our conclusions.

\section{The SIR Model}
\label{sec:sir}
 The SIR model owes its name to the three compartments that make up for the entire population, namely the group of people susceptible of getting infected, the group of infected people and the group of recovered at any given time. If we assume no mortality and a constant population $N$\footnote{More involved versions of the SIR model might allow for natality and mortality rates independent of the infection under consideration as well as mortality rates due to the infection itself, for a more detailed version discussion of these issues, the reader is refered to chapter 3 of  \cite{29}}, we can normalize everything to the value of $N$ and work with fractions, such that the variables susceptible $\cS(t)$, infected $\cI(t)$ and recovered $\cR(t)$ take values between 0 and 1. Hence, at any given time $t$ it must hold that $\cS(t)+\cI(t)+\cR(t)=1$. If we want to know the net number of infected people at any given time, we just have to multiply $\cI(t)$ by $N$ and similarly for the other variables in the model.
 
The equations governing the evolution of the simple SIR model we have just described are the following 
     \begin{eqnarray}
   \frac{d \cS}{d t}&=&-\beta \cS \cI, \\
      \frac{d \cI}{d t}&=& \beta \cS \cI-\gamma \cI, \label{eq:I}  \\
         \frac{d \cR}{d t}&=&\gamma \cI. 
   \end{eqnarray}
The model is specified in terms of the parameters $\beta$ (the infection or contagion rate) and $\gamma$ (the recovery rate). The units of the constants
are given by $[\beta]=[\gamma]=1/T$, and the fractional populations are dimensionless  $[\cS]=[\cI]=[\cR]=1$. $T$ represents the unit of time, which is considered to be a day. This system of equations appears to be independent of $N$. However, we must recall that this is a non-linear system of ordinary differential equations, the effect of $N$ can indeed be seen in the boundary conditions, for instance, if at $t=0$ we have the first reported infected person, the boundary conditions needed to evolve the SIR model of interest are
\begin{equation}\label{bc1}
\cS(0)=\frac{N-1}{N}\,,\quad \cI(0)=\frac{1}{N}\,\quad {\rm and}\quad \cR(0)=0\,. 
\end{equation}
As $N$ increases its effect on the evolution of a SIR model is going to be a delay in the appearance of the infection peak. 
More in general, one can consider to run a model in such a way that at $t=0$ there is a fraction of infected $i$ and already a fraction of recovered $r$. In this fashion, the boundary conditions will be:
\begin{equation}\label{bc11}
\cS(0)=1-i-r\,,\quad \cI(0)=i\,\quad {\rm and}\quad \cR(0)=r.\,
\end{equation}
This type of conditions are not necessarily accounted for by the solutions satisfying \eqref{bc1} and this is a natural consequence of the non-linearity of the system of differential equations. The boundary conditions \eqref{bc1} become specially relevant if one attempts to solve the SIR model by patches, where each patch has a distinct $\beta$ parameter.
It is also useful to write them in this way for the case when $N$ is a large quantity and the
SIR equations become linear. For this case the condition $\cS\approx 1$ holds,
and the SIR equations evolution is independent of a rescaling of $N$. We elaborate more on this issue when discussing time dependent SIR models in subsections \ref{s32} and \ref{s33}. 

The cumulative number of infections can be defined as
   \begin{equation}
   \cI_c(t)=\cI(t)+\cR(t)\,,
   \end{equation}
and would correspond to the total number of infections reported e.g. by John Hopkins Data Base\cite{11}, once we multiply $\cI_c(t)$ by the population $N$. 
   
At the begining of the outbreak $\cS(t)$ is roughly one, hence Eq. \eqref{eq:I} admits an approximate solution of the form: 
    \begin{equation}
\cI(t)=\cI_0\,e^{(\beta-\gamma)t}\,, \label{I0}
    \end{equation}
with this expression we can obtain the $\cR(t)$ for the exponential phase as 
    \begin{equation}
\cR(t)=\frac{\gamma \cI_0}{\beta-\gamma}e^{(\beta-\gamma)\,t}\,,\label{R00}
    \end{equation}
we can define $R_0=\beta/\gamma$ as the reproduction number, i.e. the number that
characterizes the evolution of the infection. Now if we consider the cumulative number of infections in the SIR model during the exponential phase we obtain 
\begin{equation}
\cI_c(t)=\cI(t)+\cR(t)=\frac{R_0\cI_0}{R_0-1}e^{(\beta-\gamma)\,t}.\label{Ic}
\end{equation}
\begin{figure}[h]
\centering
\includegraphics[scale=.8]{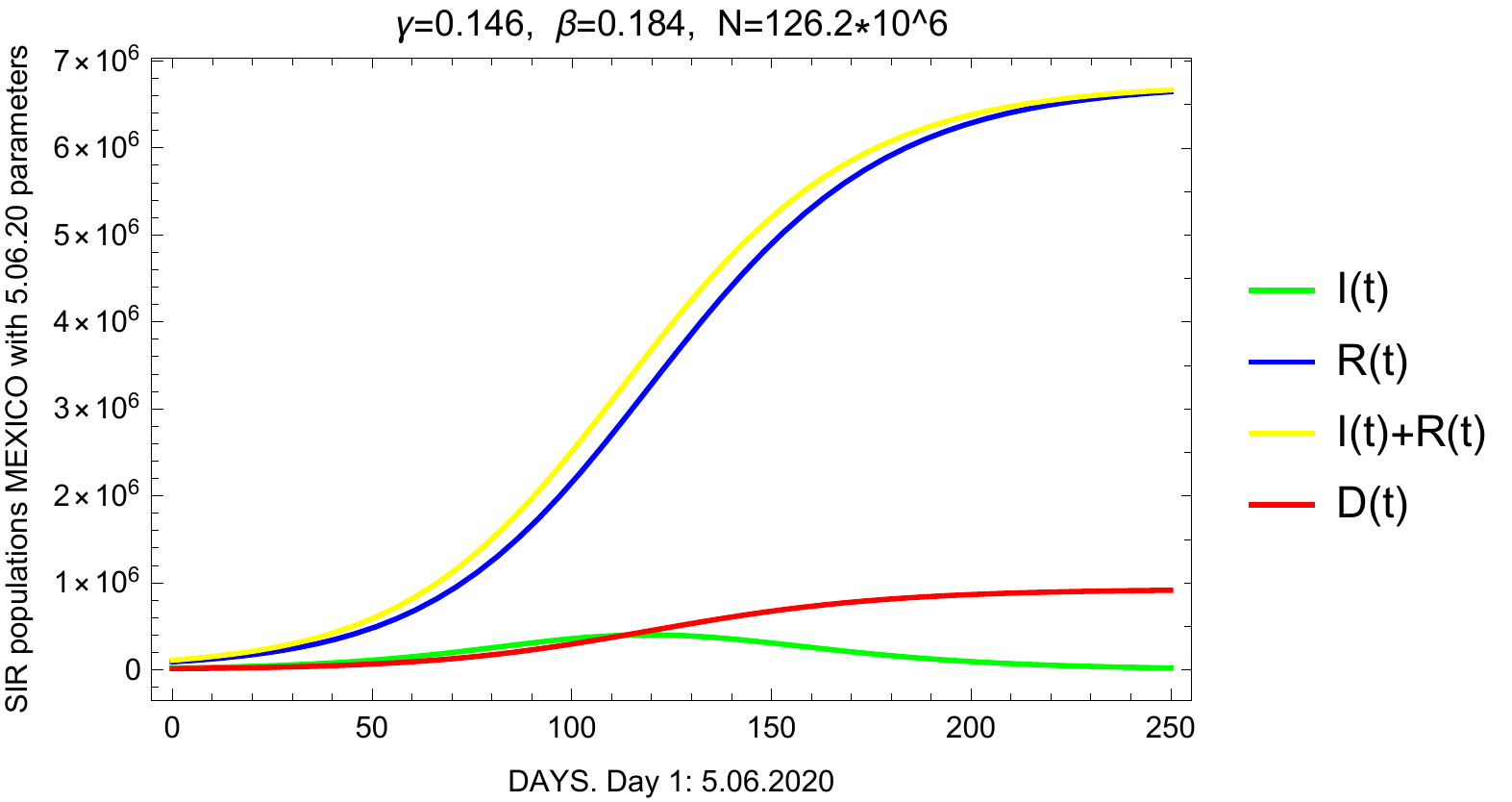}
\caption{\small{\textit{Numerical solution for the various populations of the SIR model: active cases $I(t)$ (green), recovered $R(t)$ (blue), cumulative cases $C(t)$ (yellow) and deaths $D(t)$(red). For this plot we have used  $\gamma=0.146$, $\beta=0.184$ and $N=126.2*10^6$. This
is a SIR with the population of Mexico, the contagion rate $\beta$ determined locally on $5.06.20$,
and the recovery rate determined in the interval 16.03.20 to 5.06.20. A detection rate $\alpha=0.14$
has been assumed. The deaths are estimated fitting the mortality rate $\mu=0.138$,  with $D(t)=\mu C(t)$. 
This is a worst
case scenario, which we consider it will not be reached if the $\beta(t)$
continues the current decaying dependence we explore in this work. }} }
\label{fig:Pic}
\vspace{-10pt}
\end{figure}

Our intention is to use the data of confirmed, recovered and deceased patients. Particularly for the former two, the accesible data corresponds only to a fraction of the total infected and recovered patients. One expects that under adequate testing protocols, the numbers reported keep a time independent proportionality to the actual numbers. This is reflected in the following equation
    \begin{eqnarray}
   \cI_c(t)=\alpha \cI_{TOT}(t),
    \end{eqnarray}
where $\cI_{TOT}(t)$ is the total number of cumulative cases at time $t$. The constant of proportionality is the detection rate and we denote it by $\alpha$. For instance, in China, it is considered that only 14\% of all cases were detected [5], that is $\alpha=0.14$. There are also other scenarios for which $\alpha=0.535$ [6]. 

As a final remark we must recall that our simple SIR model does not have a compartment for the deceased $\cD(t)$. The simplest manner to account for this group is to describe them as a fraction of the SIR recovered $\cR(t)$:
 \begin{equation}
 \cD(t)=\mu \cR(t)\,,
 \end{equation}
 where the a-dimensional coefficient $\mu$ corresponds to the mortality rate.  In Figure \ref{fig:Pic} we show as an example
 a global SIR model evolution for Mexico, notice the huge numbers obtained for the population.
 This is a worst case scenario that we don't expect to reach.
 
\subsection{Relation to the Logistic Sigmoid Curve}
\label{s21}
A good approximation for the number of recovered patients can be the so called Logistic Sigmoid curve, given by the following expression
\begin{equation}
\cR(t)=\frac{\cR^{(s)}}{1+e^{-\kappa (t-t_0)}},
\end{equation}
where $t=t_0$ is the inflection point, i.e. the point where the derivative attains a maximum. Note that the recovered population gets saturated at the value $\cR^{(s)}$, and that at $\cR(t_0)=\cR^{(s)}/2$. Note also that for $t\ll t_0$ the behaviour of $\cR(t)$ is essentially described by an exponential function
\begin{equation}
\cR(t)=\cR^{(s)}\,e^{\kappa (t-t_0)}+\ldots\,,\quad t\ll t_0\,,
\end{equation}
recall that $\cR(t)$ in the SIR model also has an exponential phase at the begining of the outbreak (see Eq. \eqref{R00}), from this we find that 
\begin{equation}
\kappa=\beta-\gamma\,.\label{un}
\end{equation}
In a similar fashion, it is possible to approximate the number of cumulative reported infections predicted by the SIR model in terms of a Logistic Sigmoid 
\begin{equation}
\cI_c(t)=\frac{\cI_c^{(s)}}{1+e^{-\kappa (t-t_0^\prime)}}.
\end{equation}
We have taken the same $\kappa$ since, according to the SIR model, at the beginning of the outbreak, $\cI$, $\cR$ and $\cI_c$ grow with the same exponential factor. However, we take a different inflection time $t_0^\prime$, as it can be seen from figure \ref{fig:Pic} that there is a certain offset between $\cR$ and $\cI_c$. Comparing $\cI_c$ in the exponential phase with Eq.  \eqref{Ic} we obtain 
\begin{equation}
\cI_c^{(s)}\,e^{-\kappa t_0^\prime}=\frac{R_0 \cI_0}{R_0-1}\,,\label{dos}
\end{equation}
using Eqs. \eqref{un} and \eqref{dos} we obtain 
\begin{equation}
\cI_c^{(s)}e^{-\kappa t_0^\prime}=\tilde{R}_0 \cR^{(s)} e^{-\kappa t_0}\,.
\end{equation}
Let us recall that for large $t$, the quantity $\cI(t)$ in the SIR model approaches zero. Hence, the asymptotic values of $\cI_c^{(s)}=\cR^{(s)}$ must coincide. From this observation we arrive at the following relation
\begin{equation}
e^{\kappa(t_0-t_0^\prime)}=R_0\,.
\end{equation}

 \subsection{On the Basic Reproduction Number $R_0$}
 \label{s22}
For the SIR model with constant parameters one can establish time dependent quantities in order to account for the evolution of the epidemic. One can draw some inspiration from the Logistic curve approximation and define 
\begin{equation}
\kappa^{(1)}(t)=\frac{d\,{\rm Log}(\cI_{c}(t))}{dt}.
\end{equation}
Note then that at the beginning of the outbreak, $\kappa(t)=\kappa$ as defined in Equation \eqref{un}. In this fashion, we can define a time dependent basic reproduction number 
\begin{equation}
R_0^{(1)}(t)=1+\frac{\kappa^{(1)}(t)}{\gamma}\,,\label{R01}
\end{equation}
note that this quantity is always bigger than 1, matches the global basic reproduction number at the beginning of the outbreak and as it approaches 1 it signals the end of the disease's spread. This
definition can also be employed for a time varying $\beta(t)$ in the  limit of $N$ huge,
such that $R_0^{(2)}(t)=1+(\cI+\cR)'/(\gamma (\cI+\cR))$.

In a similar fashion we can use the infected population in order to define a second dynamical quantity, 
\begin{equation}
\kappa^{(2)}(t)=\frac{d\,{\rm Log}(\cI(t))}{dt}=\beta \cS(t)-\gamma,
\end{equation}
in this case the dynamical basic reproduction number can be defined analogously, 
\begin{equation}
R_0^{(2)}(t)=1+\frac{\kappa^{(2)}(t)}{\gamma}\,,\label{R02}
\end{equation}
Note that in this case, $R_0^{(2)}(t)$ becomes 1 once the infection peak is reaches and for further times is less than one. An $R_0^{(2)}$ smaller than one is a sign that the infected population is decreasing. 
Again this definition can be considered for a time varying $\beta(t)$ in the  limit of $N$ huge, giving $\cS(t)\approx 1$,
such that $R_0^{(2)}(t)=1+\cI'/(\gamma \cI)$.

When the contagion rate of the SIR model is dependent of time, which  is the case of the second approach considered in this work the value of
reproduction number can be defined as:
\begin{equation}
R_0^{(3)}(t)=\frac{\beta(t)}{\gamma}.
\end{equation}
This  quantity in the limit $\cS(t) \approx 1$  which is the situation in our second model, coincides with both of
the previously defined ones $R_0^{(1)}(t)$ and $R_0^{(2)}(t)$.

\section{Forecast Methodologies}
\label{s3}

In this section we describe the two methodologies applied in fitting the various models to the data. In order to do so, we must first prepare the data in order to fit the observables of the SIR model. For the data we denote $C(t)$ the number of confirmed cases for a given time (day) $t$. Similarly we denote the recovered by $R(t)$ and the deceased by $D(t)$. Note that the SIR model does not have a compartment for the deceased, hence this quantity has to be assigned to the recovered compartment. In our case we choose to combine the data for both recovered and deceased into a new ``effective recovered" group $R_{SIR}(t)$, more suited to be described by the SIR model
\begin{equation}
R_{SIR}(t)=R(t)+D(t)\,,
\end{equation}
and similarly, we describe the infected active by
\begin{equation}
I_{SIR}(t)=C(t)-R(t)-D(t)\,.
\end{equation}
The data we would like to fit is now described in terms of $C(t)$, $R_{SIR}(t)$ and $I_{SIR}(t)$. As a notation reminder, datasets we denote by italic letters, $C$, $R$, $I$, etc, whereas functions, such as the ones in the SIR model are denoted by curly letters $\mathcal{C}$, $\cS$, $\cI$, $\cR$, etc. 

\subsection{Time Independent SIR Model}
\label{s31}

For this methodology the goal is to obtain the optimal values for the population $N_{eff}$, the contagion rate $\beta$ as well as the recovery rate $\gamma$, such that the data for  confirmed, recovered and active infections is best fitted with a SIR model. Note that in the for the initial conditions presented in Eq. \ref{bc1}, Day 0, i.e. $t=0$ is the day when the first confirmed case is reported. However, one has to be careful with setting $t=0$, as the first infection might have occurred before. For that purpose we also include a time offset, or delay $\delta t$, in such a manner that the SIR model observables used to fit the data are $\cI_c(t+\delta t)$, $\cR(t+\delta t)$ and $\cI(t+\delta t)$. Therefore, in addition to the SIR parameters $N$, $\beta$ and $\gamma$ we would also have to estimate $\delta t$. As mentioned already, the data we intend to fit are the confirmed cases $C(t)$, to be described by the function $\cI_c(t+\delta t)$, the sum of recovered plus deceased $R(t)+D(t)$ to be described by the function $\cR(t+\delta t)$ as well as the active cases $C(t)-R(t)-D(t)$ to be described by $\cI(t+\delta t)$. 

Note that at the begining of the outbreak we could fit an exponential to $C(t)$ in doing so we would obtain the the parameter $\kappa=\beta-\gamma$. From this approach one would have to be careful since only as small amount of data follows a straight line in log scale. This implies that the most recent data can not be used for the estimation of $\kappa$, therefore missing the effect of the most recent interventions, nevertheless this
can serve to a time local description of $\beta(t)$ as the ones discussed later. Furthermore, in choosing a given interval we would be introducing bias errors in the estimation of $\kappa$. Instead we fit $C(t)$ by a sigmoid function and estimate $\kappa$ from the best fit. We use Mathematica for the estimation of the best fit. 

Having the value of $\kappa$ we reduce the parameter space by 1: We leave $\beta$ out of the game and estimate it using $\kappa$ once we obtain the value of $\gamma$. Note that we don't employ analytic expressions for $\cI_c(t)$, $\cR(t)$ and $\cI(t)$, for this SIR model with time delay, however an analytical expression for the standard SIR is well studied \cite{30}. We have to numerically solve the SIR differential equations for fixed $N$, $\beta$ and $\gamma$. In order to describe numerically how close a given model approaches the data we define the following likelihood function
\begin{align}
\begin{split}
L(N,\gamma,\delta t)=\sum_{t=0}^{\rm Today}&\left[(\cI_c(t+\delta t)-C(t))^2+(\cR(t+\delta t)-R_{SIR}(t))^2\right. \\
&\,\left. +(\cI(t+\delta t)-I_{SIR}(t))^2\right]\,,
\end{split}
\end{align}
the optimal values for $N$, $\gamma$ and $\delta t$ are those for which $L(N,\gamma,\delta t)$ attains a minimum. The errors are estimated from the function $L(N,\gamma,\delta t)$ as well, depending on how far one has to go in either direction such that the value of $L$ increases by 0.5 above the minimum. 

The error estimates for $N$, $\gamma$, $\delta t$ and $\beta$ (through $\kappa$) are used to run SIR models with different configurations of parameters, i.e.\footnote{In general upper and lower error bars need not to be the same.}
\begin{equation}
\{p_1=(N+\Delta N_+,\beta,\gamma,\delta t)\,,\, p_2=(N-\Delta N_-,\beta,\gamma,\delta t)\,\,p_3= (N,\beta+\Delta\beta_+,\gamma,\delta t)\,,\ldots \,,\}_I
\end{equation}
where the index $I=1,...,10$ denotes a set of particular parameter values, i.e. an element on the net of parameters.  In each of these situations we compare to the SIR model for the central value parameters $p_0=(N,\beta,\gamma,\delta t)$. Say we are interested in the prediction for cumulative infections $\cI_c$, the central value will be given by $\cI_c(p_0;t)$ and for the corresponding Error bars, we define two sets
\begin{align}
P^+(t)=&\{p_I\,|\,\cI_c(p_I;t)-\cI_c(p_I;t)>0\}\,,\\
P^-(t)=&\{p_I\,|\,\cI_c(p_I;t)-\cI_c(p_I;t)<0\}.
\end{align} 
Then the errors in $\cI_c$ at time $t$ are given by 
\begin{align}
\Delta \cI_c^+(t)=&\sqrt{\frac{1}{N_+-1}\sum_{p_I\in P^+(t)}(\cI_c(p_I;t)-\cI_c(p_0;t))^2}\,,\\
\Delta \cI_c^-(t)=&\sqrt{\frac{1}{N_--1}\sum_{p_I\in P^-(t)}(\cI_c(p_I;t)-\cI_c(p_0;t))^2},
\end{align} 
where $N_+$ and $N_-$ are the number of elements in $P^+(t)$ and $P^-(t)$ (usually 4). The confidence interval for $\cI_c$ at time $t$ lies then between $\cI_c(p;t)-\Delta \cI_c^-(t)$ and $\cI_c(p;t)+\Delta \cI_c^+(t)$. 

The optimal values and the corresponding errors for the parameters pertaining each country are reported in Table \ref{tab:PD}.

\begin{table}[h]
\centering
\renewcommand{\arraystretch}{1.4}
{\small
\begin{tabular}{|c|c|c|c|c|c|}\cline{2-6}
\multicolumn{1}{c|}{} & $\beta$ & $\gamma$& $N_{eff}$ & $\mu$ & $\delta t$ \\ \hline
ARGENTINA & $0.078\pm 0.008$  & $0.025\pm 0.007$ & $(1.20^{+3}_{-0.4})\cdot\,10^5$ & $0.15\pm0.09$ & $50\pm 4$ \\ \hline
BRAZIL & $0.124\pm 0.007$  & $0.052\pm 0.005$ & $(2.1^{+0.4}_{-0.2})\cdot\,10^6$ & $0.13\pm0.01$ & $50\pm 3$\\ \hline
COLOMBIA & $0.078\pm 0.007$  & $0.022\pm 0.007$ & $(1.3^{+0.9}_{-0.5})\cdot\,10^5$ & $0.12\pm0.03$ & $47^{+5}_{-4}$\\ \hline
MEXICO & $0.24\pm 0.05$  & $0.18\pm 0.05$ & $(5.5\pm 1)\cdot\,10^5$ & $0.14\pm0.02$ & $31\pm 3$\\ \hline
SOUTH AFRICA & $0.12\pm 0.01$  & $0.06\pm 0.01$ & $(6^{+5}_{-3})\cdot\,10^5$ & $0.05\pm0.03$ & $39\pm 4$\\ \hline
\end{tabular}
}
\caption{\textit{Best fit parameters for the time independent, and with time-delayed SIR model and their corresponding errors. ($\delta t$ is given in days).)}} 
\label{tab:PD}
\end{table} 

The figures \ref{AR}, \ref{BR}, \ref{CO}, \ref{MX} and \ref{ZA} show in red lines the forecast of this model for
Argentina, Brazil, Colombia, Mexico and South Africa respectively. In each figure the 1st, 2nd, 3rd and 4rd plots
show: the cumulative cases $\mathcal{C}(t)$, the deaths $\cD(t)$, the active cases $\cI(t)$ and the values of $R_0$. The confidence intervals are given in red shadowed regions. The same figures show for contrast the other two models discussed in the
next subsections.

\subsection{Step changing contagion rate}
\label{s32}

In this subsection we implement a SIR model with a time changing contagion rate in order to describe the evolution of the various population compartments for the considered countries.  The first approach is to study $\beta(t)$ as a step function, where each
time interval has constant values of it. In order to extrapolate
we consider a functional dependence for $\beta(t)$, which is chosen to be 
an exponential decay \cite{26}.  Throughout this section the total population 
$N$ is taken to be the entire population of each country. We also estimate
the recovery rate by fitting the experimental data $\frac{dR(t)}{dt}$ vs. $I_{SIR}(t)$ from the SIR.

From the available data we set the contagion rate change in discrete time steps. 
For this purpose we divide the data in intervals between 
 4 and 6 days, depending on which choice does a better fitting to 
 the data of a given country. The decrease in $\beta$ as time evolves
obeys to the country contention measures, and it doesn't assume
a particular dependence. Therefore it describes very well the measurements.

In this section we implement a SIR model with a time changing contagion rate in order to describe the evolution of the various population compartments for the considered countries.   In the first place we take $\beta(t)$ as a step function. We  adjust local exponentials to (\ref{R01}) and (\ref{R02}) to every region of constant $\beta(t)$.  In the
next subsection \ref{s33} we consider an exponential dependence, which allows to extrapolate
to the future.   Through out the subsection the total population 
before the outbreak is considered the country population $N$. We also estimate
the recovery rate by fitting the experimental data $\frac{dR}{dt}$ vs. $I_{SIR}$, notice
than here $R(t)$ has no subindices, meaning that this is the real data of recovered.
It is important to say, that a similar study with the fit of $\frac{dR_{SIR}}{dt}$ vs. $I_{SIR}$,
as the one performed in previous section could improve the results for the deaths estimates, 
however the parameter $\gamma$ obtained has a different interpretation than the real recovery rate.



Let us start by explaining the procedure. For all the countries studied we consider 
that the total population $N$ is the number of inhabitants of the country. The contention measurements are
then only reflected on the change of the contagion rate $\beta(t)$. For this case if no herd
immunity is pursued, as  $\mathcal{S}\approx 1$ we are in the linear regime of the differential equations \cite{26}.  
Meaning that the solutions of the $\cI$, $\cR$ and $(\cI+\cR)$ obey an exponential dependence. This dependence can be
explored at any time of the evolution and in the used units gives the results in (\ref{R00}) and (\ref{I0}).
Then, if after the contention measures are implemented $\beta(t)<\gamma$ is achieved, only a small fraction of the population 
will be infected.

There are two important points to discuss about the linearity of the equations in this regime of large $N$. One was noted in \cite{26}, this is that assuming a constant in time detection rate ($\alpha$) such that the real population group numbers are denoted
by subindex $TOT$, and the detected quantities are given by $\mathcal{I}=\alpha\cI_{TOT},  \mathcal{R}=\alpha\cR_{TOT}$, 
the evolution of $\cI, \cR$ does not change for different values of $\alpha$. This point is central in our analysis.
\footnote{Is important to note that for the non linear regime there is a big difference of the SIR
evolution for different values of $\alpha$ \cite{26}.}
This is due to the fact that the equations are linear and the initial conditions
for infected and recovered are also obtained by multiplication $\mathcal{\cI}_0=\alpha I_0,  \mathcal{\cR}_0=\alpha R_0$. 
Such that one can solve only for the detected quantities ($\cI,\cR)$.
Analogously if the population number $N$ is so big such that $\cS \approx 1$ the differential equations evolution is independent
on the actual value of $N$. This is again due to the linearity of the equations
and of the initial conditions. We have thus solved numerically the  SIR equations for a $\beta(t)$ varying in time:
     \begin{eqnarray}
     \frac{d \cS_{TOT}}{d t}&=& \beta \cS_{TOT}\cI_{TOT},  \, \,  \,  \,  \, \frac{d\cI_{TOT}}{d t}= (\beta \cS_{TOT}-\gamma)\cI_{TOT},  \, \, \,  \,  \,  \,     \frac{d\cR_{TOT}}{d t}=\gamma\cI_{TOT}, \label{eqL}\\
        \cI_{TOT}(0)&=&\frac{I_0}{\alpha N}, \,   \,  \,   \,  \cR_{TOT}(0)=\frac{R_0}{\alpha N},  \,  \,   \,  \,   \cS_{TOT}(0)=1-\frac{I_0+R_0}{\alpha N}, \nonumber
   \end{eqnarray}
The initial conditions consider that at time $t=0$ there are $I_0$ detected infected persons, 
and $R_0$ detected recovered persons. The solutions of (\ref{eqL}) in linear regime reduce to solutions of the system:
     \begin{eqnarray}
     \cS_{TOT}&\approx& 1,  \frac{d \cI_{TOT}}{d t}= (\beta-\gamma) \cI_{TOT},  \,     \,   \,    \,   \,    \frac{d \cR_{TOT}}{d t}=\gamma \cI_{TOT}, \\
         \cI_{TOT}(0)&=&\frac{I_0}{\alpha N}, \, \cR_{TOT}(0)=\frac{R_0}{\alpha N},  \cS_{TOT}(0) \approx 1. \nonumber
   \end{eqnarray}
As the equations are linear the substitutions $\cI_{TOT}\rightarrow  \cI /\lambda$,  $\cR_{TOT} \rightarrow  \cR/\lambda$ just change the initial
conditions to: $\cI (0)=\frac{I_0 \lambda}{\alpha N}=\frac{I_0 }{N'}$, $\cR(0)=\frac{R_0 \lambda}{\alpha N}=\frac{R_0}{N'}$.  This illustrates the fact
that the predictions are independent of $N$.  Now let us consider just $N'=N$, then for the detected populations we have the system:
     \begin{eqnarray}
     \cS&\approx& 1,  \frac{d \cI}{d t}= (\beta-\gamma) \cI,  \,     \,   \,    \,   \,    \frac{d \cR}{d t}=\gamma \cI,  \label{linEQ}\\
         \cI(0)&=&\frac{I_0}{N}, \, \cR(0)=\frac{R_0}{N},  \cS(0) \approx 1. \nonumber
   \end{eqnarray}

Therefore in the linear regime, for a local region in time,  our obtained results are exponentials and are 
approximately independent of $\alpha$ and $N$. The detected populations of recovered and infected persons can be written as: 
 \begin{eqnarray}
N \cI &=&   {I}_0 \text{exp}((\beta-\gamma) t), \label{eq:I3}  \\
N  \cR&=&  ({R}_0-\delta) \text{exp}((\beta-\gamma) t)+\delta,   \nonumber \\
 &=&  \frac{\gamma {I}_0}{\beta-\gamma} \text{exp}((\beta-\gamma) t)+\delta.    \nonumber
   \end{eqnarray}

To study the evolution of the systems we first fit the values of $\gamma$
by making a min-square fit of the linear dependence $\frac{d R}{d t}$ versus
$I_{SIR}$ in (\ref{linEQ}). There is some universality in these values, however there 
are differences between the countries. Such differences can be coding the fact that
the recovered data are reported in different ways.
In the Table \ref{tab:PN} we summarize the $\gamma$ of the five cases
studied with date till $05.06.20$.

\begin{table}[h]
\centering
\renewcommand{\arraystretch}{1.4}
{\small
\begin{tabular}{|c|c|c|c|c|c|}\cline{2-6}
\multicolumn{1}{c|}{} & \multicolumn{2}{c|}{$R_0^{(1)}$}  & \multicolumn{2}{c|}{$R_0^{(2)}$}   & \multirow{2}{*}{$\gamma$} \\ \cline{2-5}
\multicolumn{1}{c|}{} & $\beta_0$ & $b_0$ & $\beta_0$ & $b_0$ & \\ \hline
ARGENTINA & $0.21\pm0.03$& $0.023\pm0.006$ & $0.19\pm0.04$& $0.021\pm0.008$& $ 0.030\pm0.008$  \\ \hline
BRAZIL & $0.26\pm0.03$ & $0.015\pm0.003$ & $0.23\pm0.07$ & $0.015\pm0.009$ & $0.064\pm 0.007$ \\ \hline
COLOMBIA & $0.22\pm0.03$ &$0.017\pm0.004$ &  $0.22\pm0.03$& $0.019\pm0.004$ & $0.045\pm0.007$ \\ \hline
MEXICO &  $0.32\pm0.01$ & $0.008\pm0.002$ & $0.30\pm0.08$ & $0.010\pm0.007$ & $0.15\pm0.02$ \\ \hline
S. AFRICA & $0.25\pm 0.04$ & $0.009\pm0.004$ & $0.24\pm0.05$ &$0.01\pm0.006$ & $0.096\pm 0.012$  \\ \hline
\end{tabular}
}
\caption{\textit{Contagion rate parameters for an exponential $\beta(t)$ and recovery rate $\gamma$ with 
confidence intervals for various countries. To obtain $\beta(t)$ the mean values
of $\gamma$ shown are the ones considered. The data
are obtained from 16.03.20 to 5.06.20.  }} 
\label{tab:PN}
\end{table}

\begin{figure}[h]
\begin{center}
\includegraphics[width=.48\textwidth]{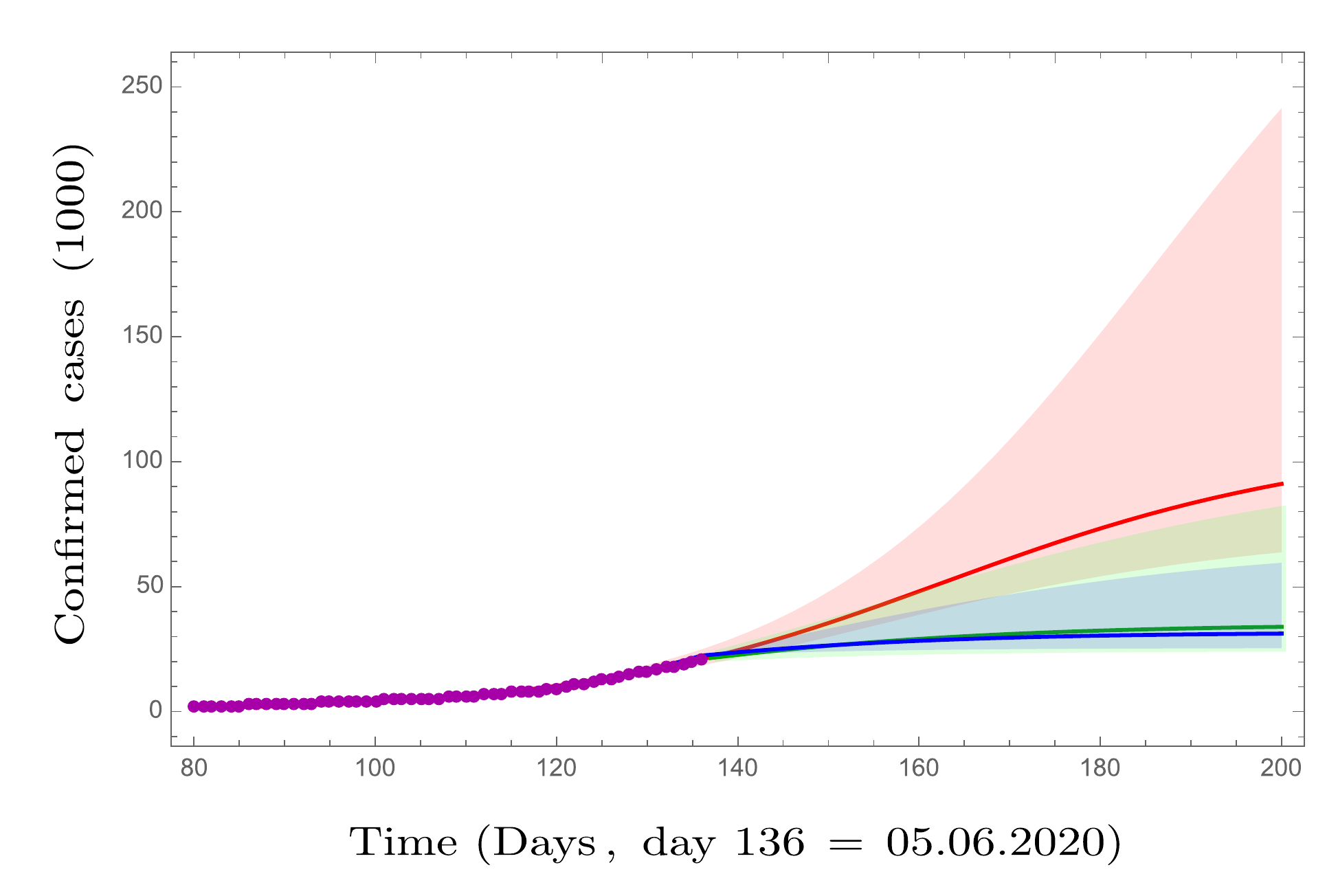}   
\includegraphics[width=.48\textwidth]{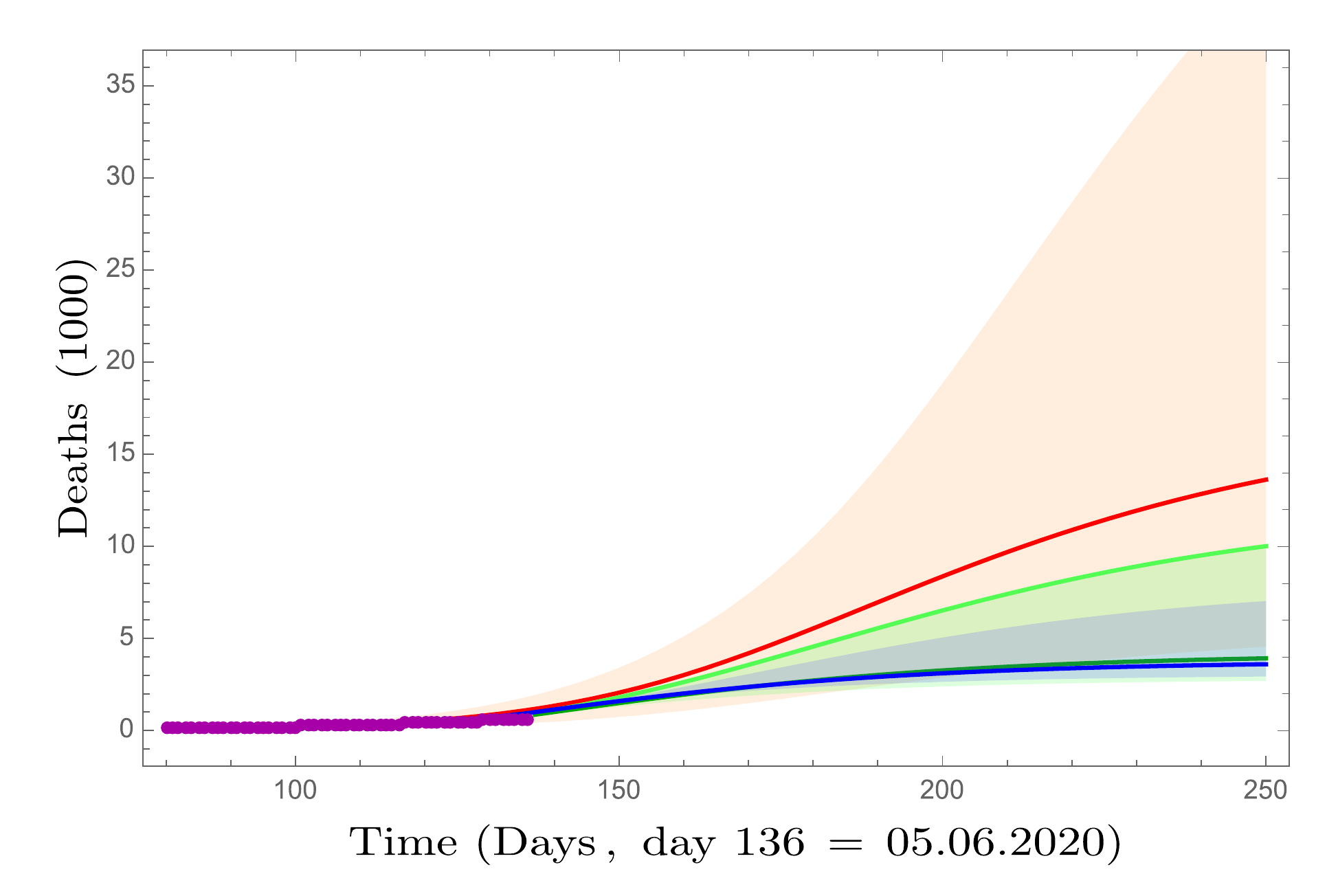}   
\includegraphics[width=.48\textwidth]{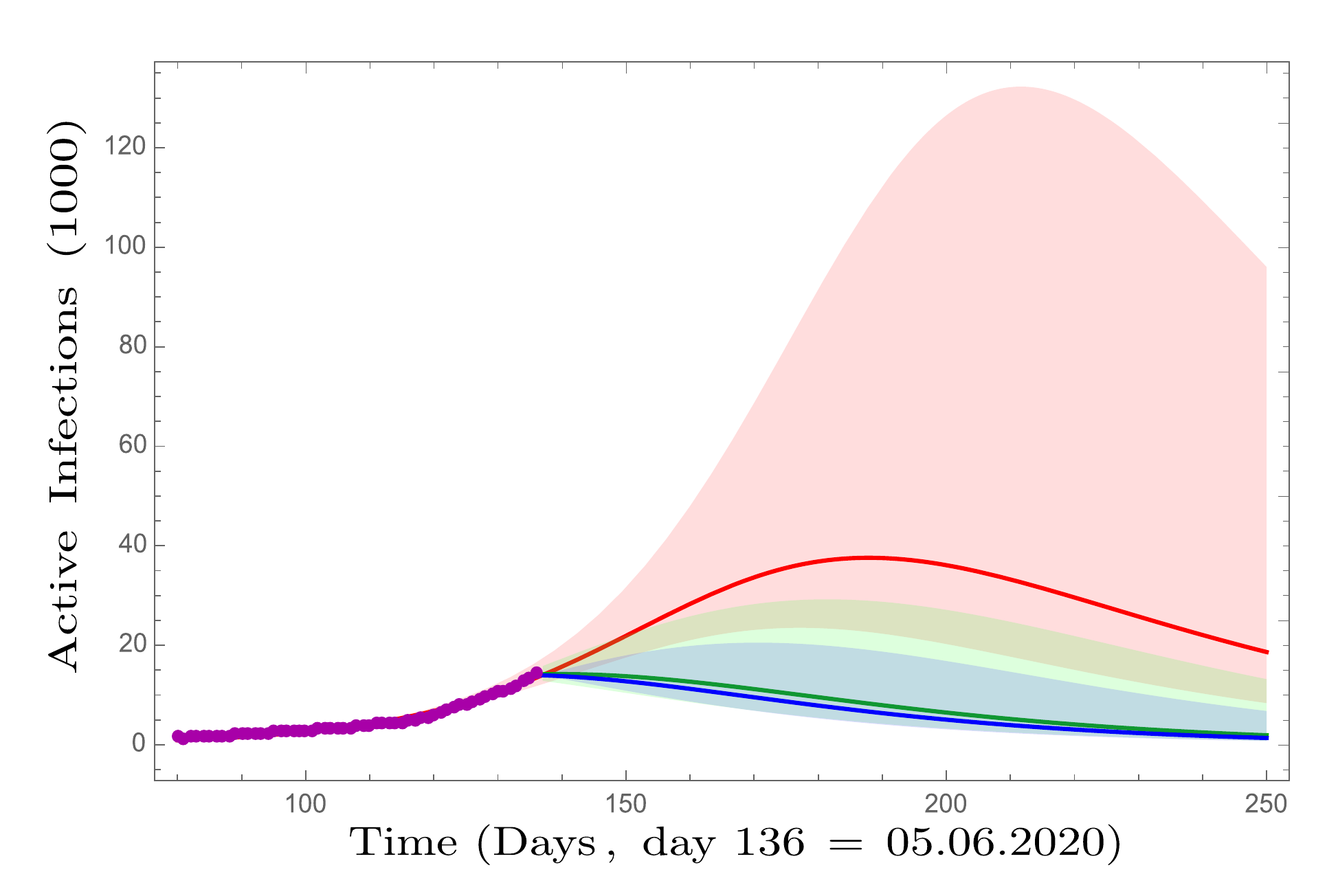}
\includegraphics[width=.48\textwidth]{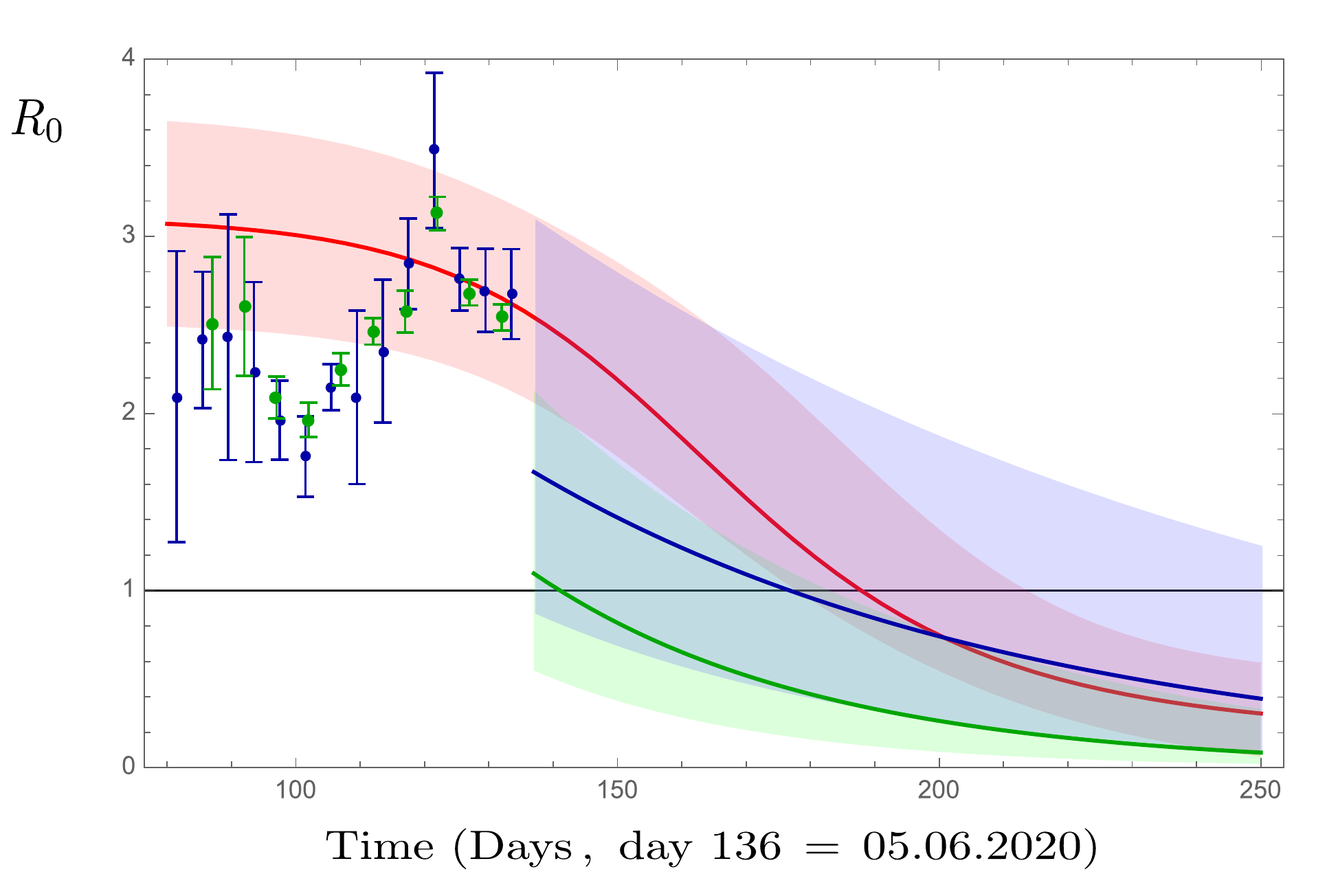}
\caption{\emph{Confirmed cases, deaths, active infections and $R_0$ versus time for Argentina
with data until $05.06.2020$. The red curves and error bars correspond to the prediction of the time independent SIR model. 
The green curves correspond to the prediction of a SIR model with time
dependent $\beta$ estimated from $(\gamma+I'/I)$ first by steps, and form an exponential fit. The blue curves correspond to the prediction of a SIR model with time
dependent $\beta$ estimated from $(\gamma+(I+R)'/(I+R))$ as an exponential fit. The error bars give a in interval with $95\%$ confidence.}}
\label{AR}
\end{center}
\end{figure}

\begin{figure}[h]
\begin{center}
\includegraphics[width=.48\textwidth]{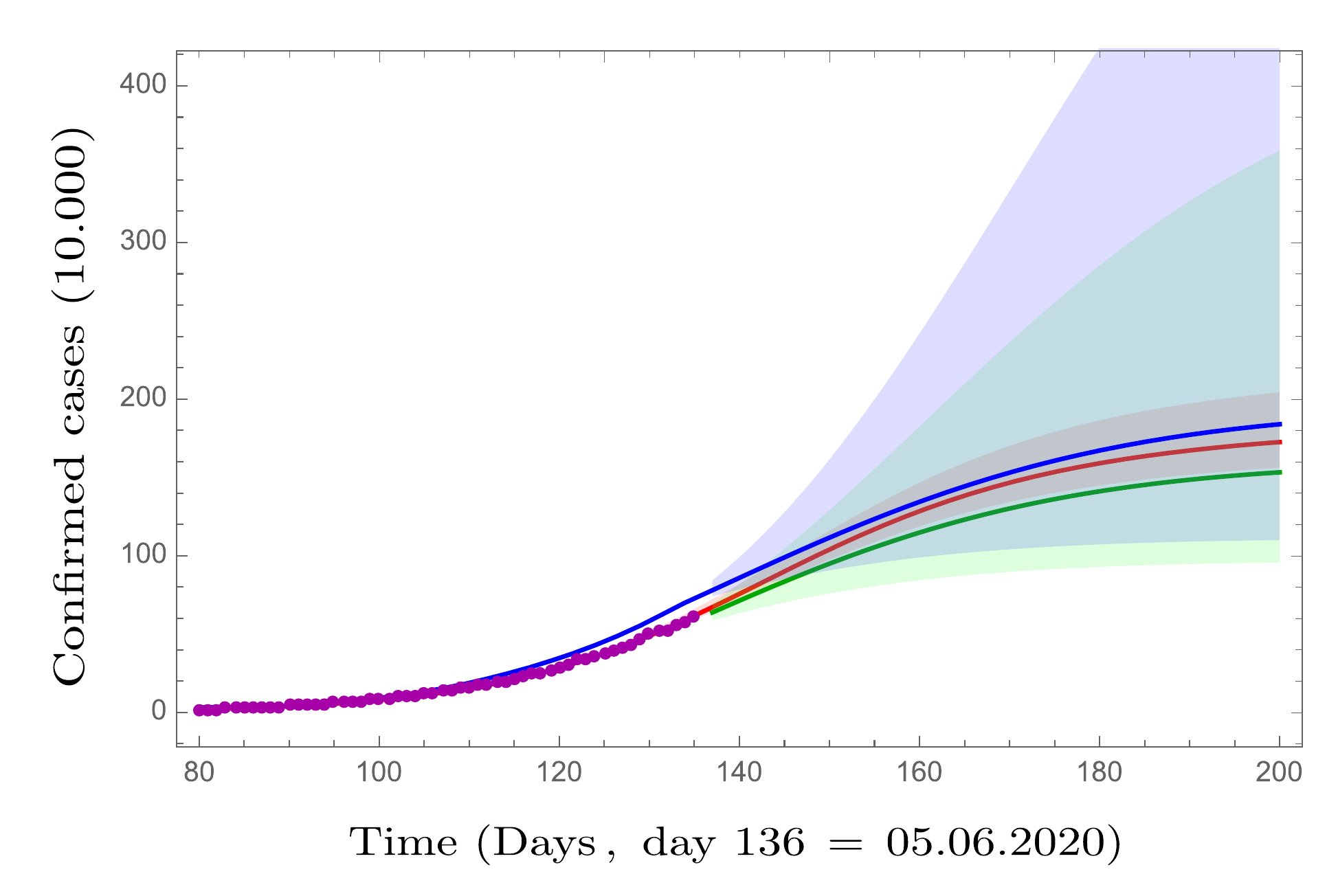}   
\includegraphics[width=.48\textwidth]{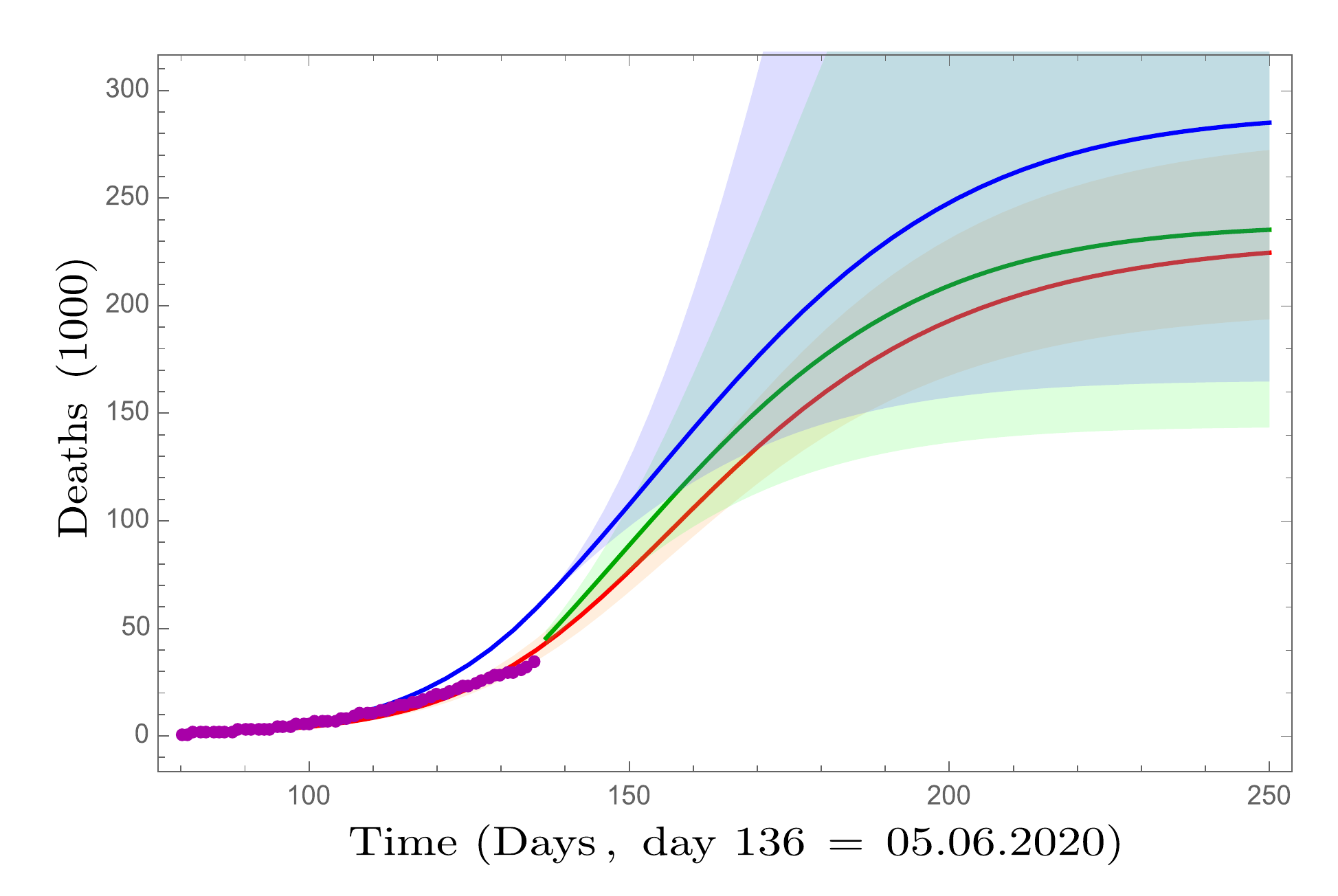}   
\includegraphics[width=.48\textwidth]{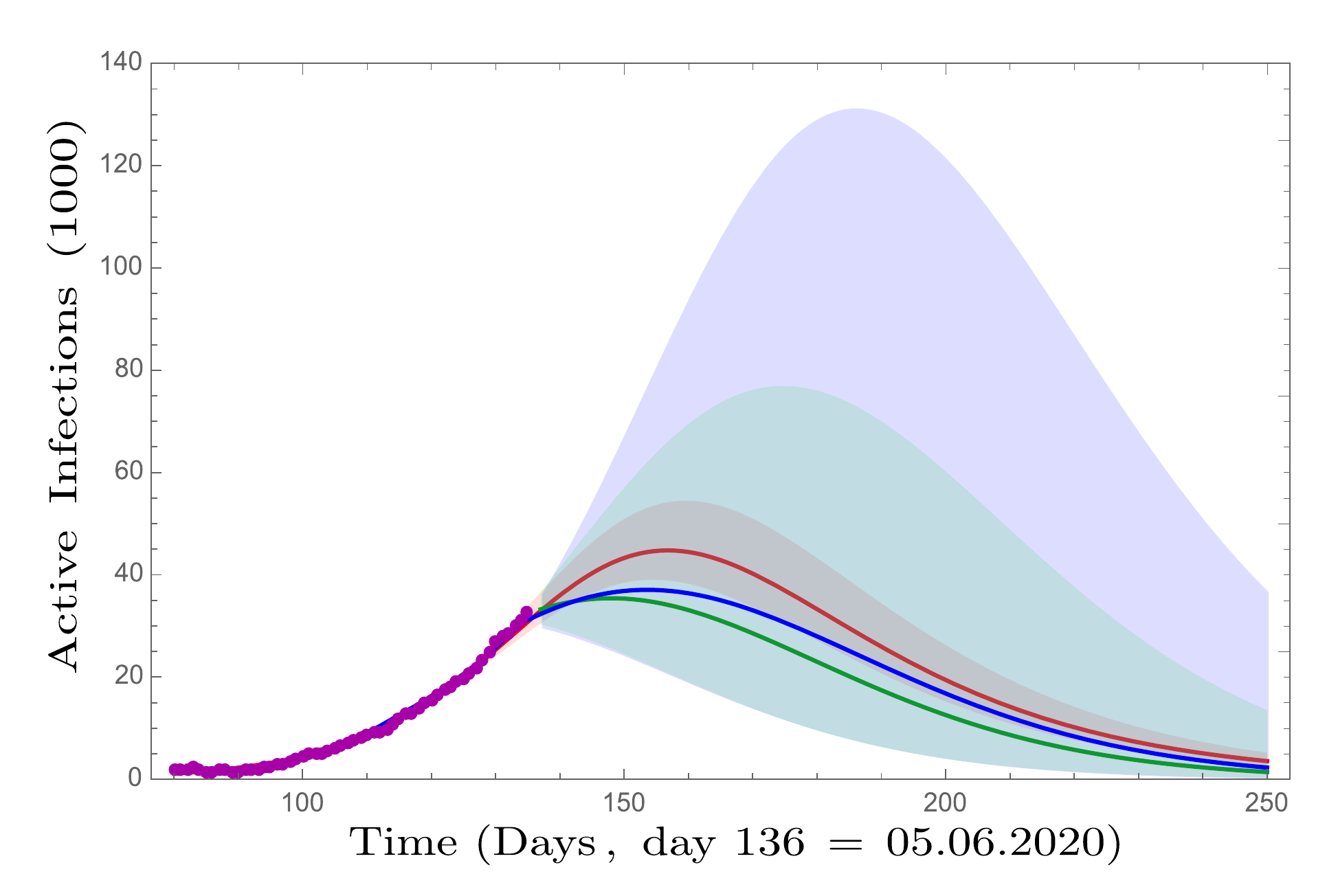}
\includegraphics[width=.48\textwidth]{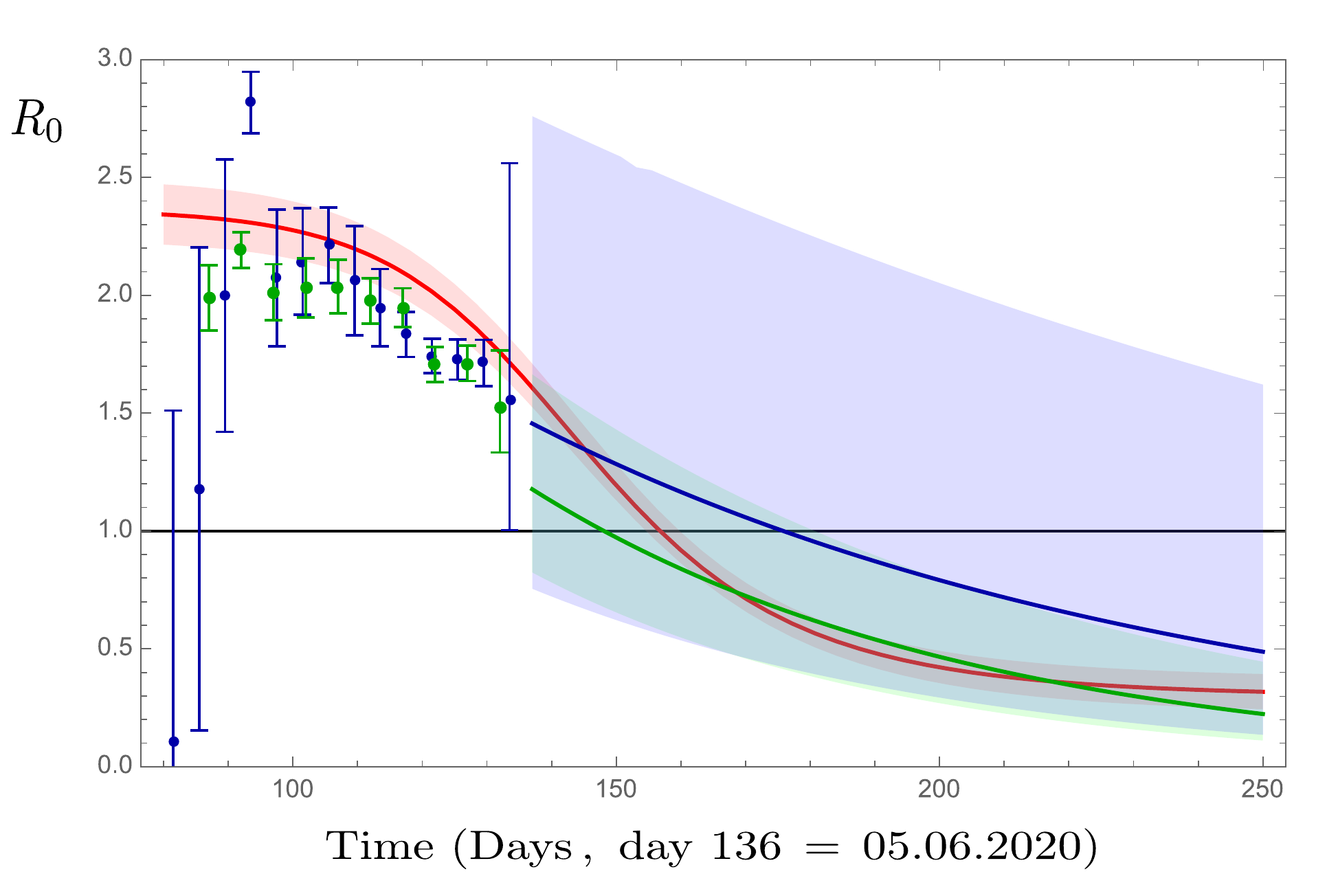}
\caption{\emph{Confirmed cases, deaths, active infections and $R_0$ versus time for Brazil
with data until $05.06.2020$. The red curves and error bars correspond to the prediction of the time independent SIR model. The green curves correspond to the prediction of a SIR model with time
dependent $\beta$ estimated from $(\gamma+I'/I)$ first by steps, and form an exponential fit. The blue curves correspond to the prediction of a SIR model with time
dependent $\beta$ estimated from $(\gamma+(I+R)'/(I+R))$ as an exponential fit. The error bars give a in interval with $95\%$ confidence
interval.}}
\label{BR}
\end{center}
\end{figure}

\begin{figure}[h]
\begin{center}
\includegraphics[width=.48\textwidth]{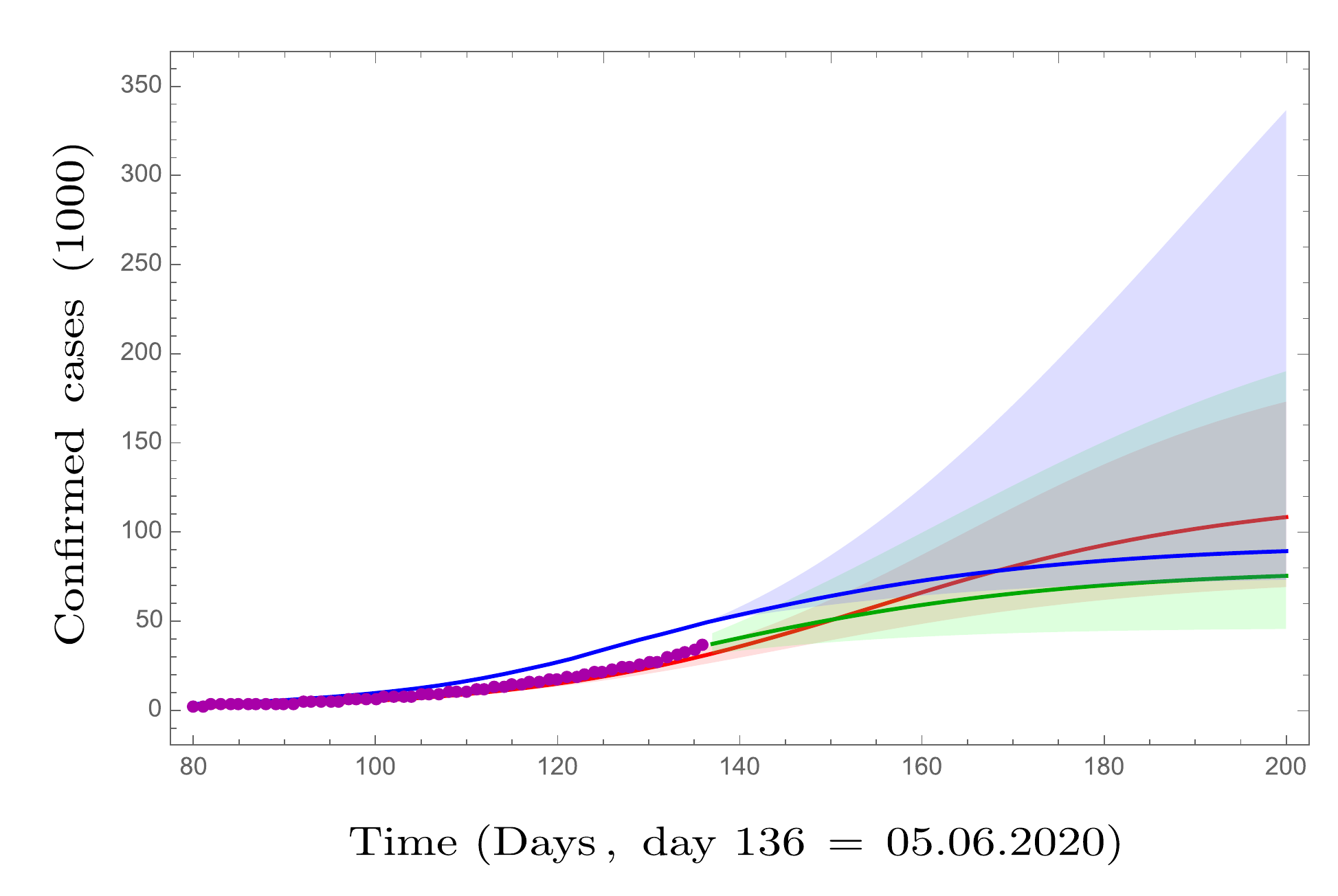}   
\includegraphics[width=.48\textwidth]{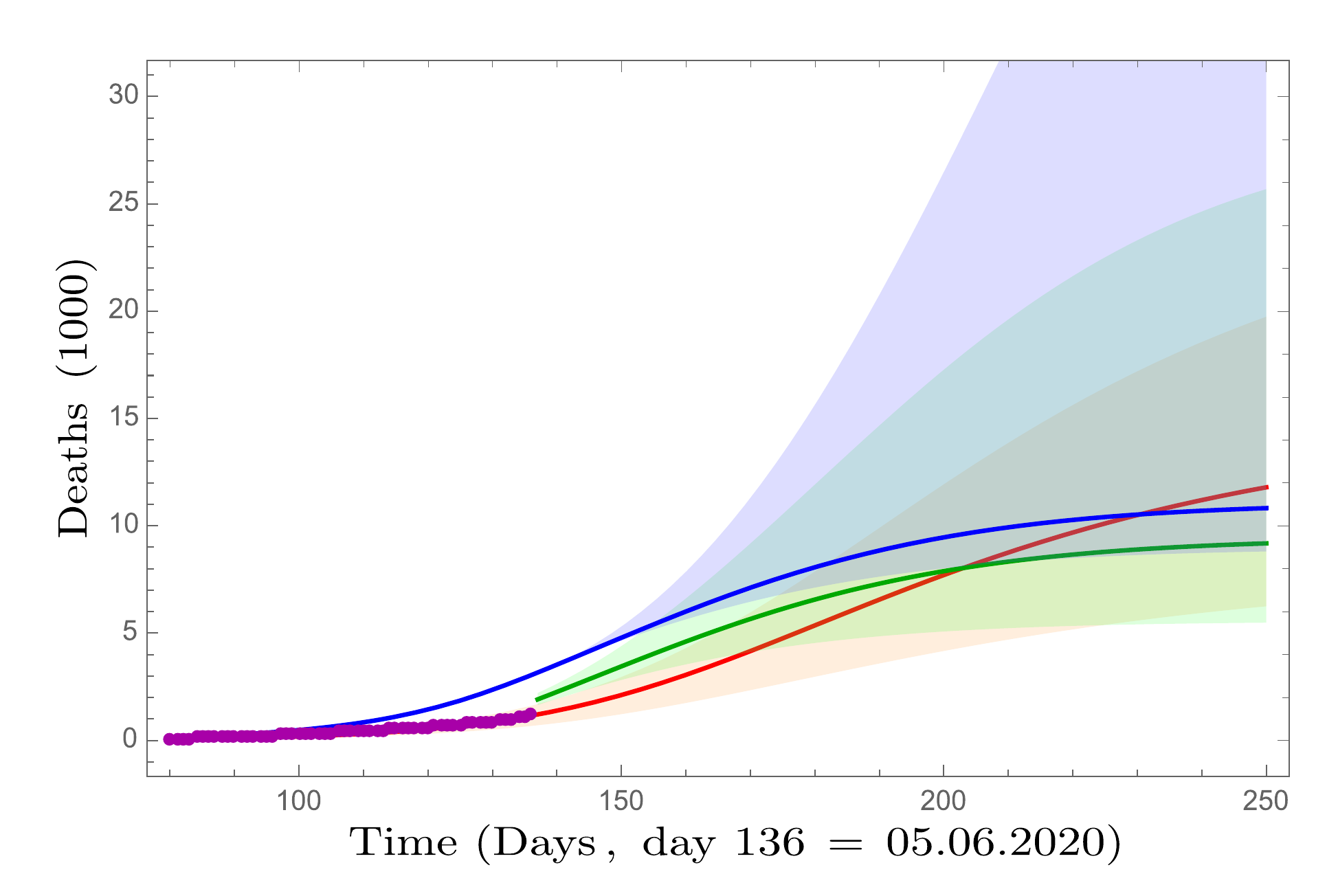}   
\includegraphics[width=.48\textwidth]{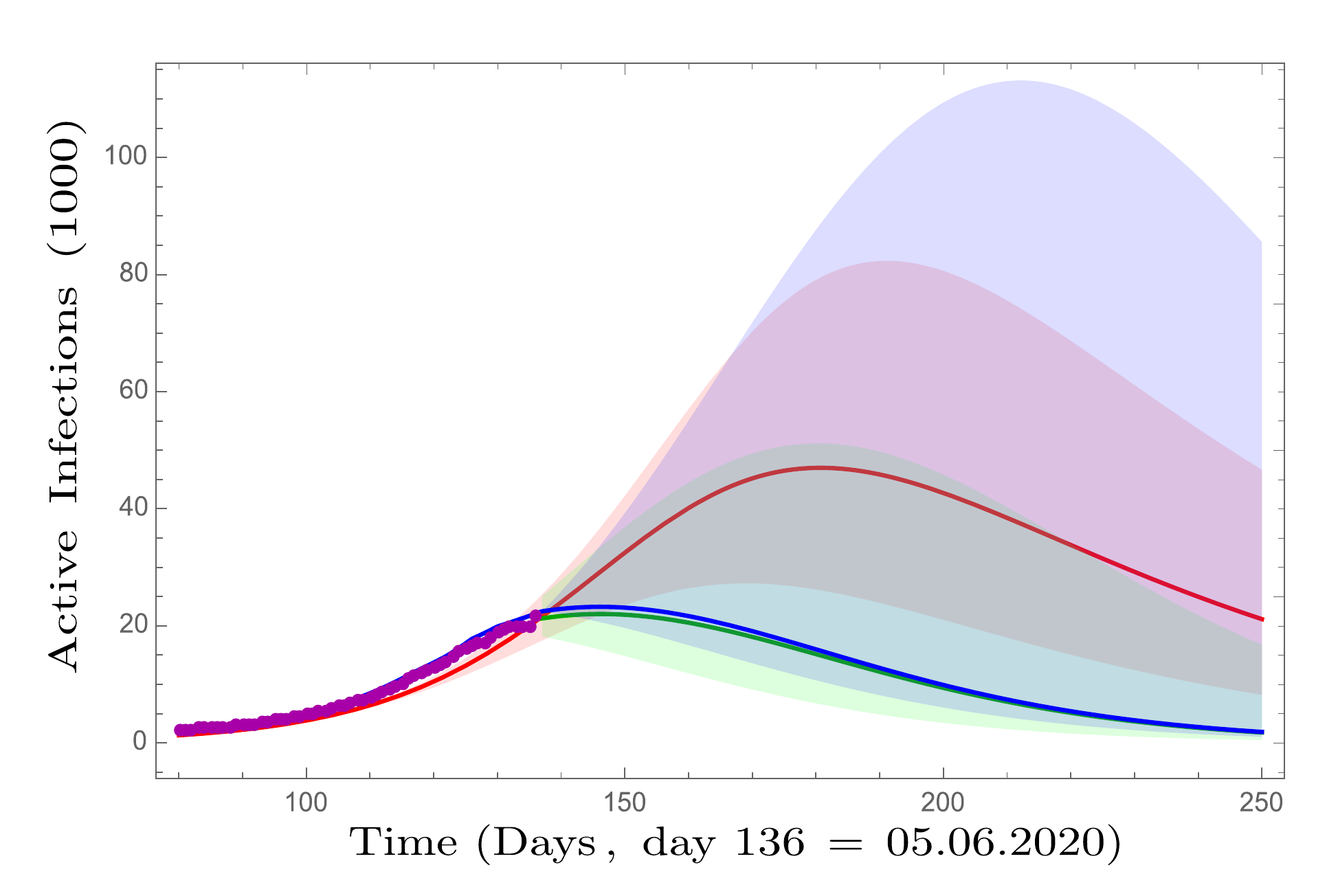}
\includegraphics[width=.48\textwidth]{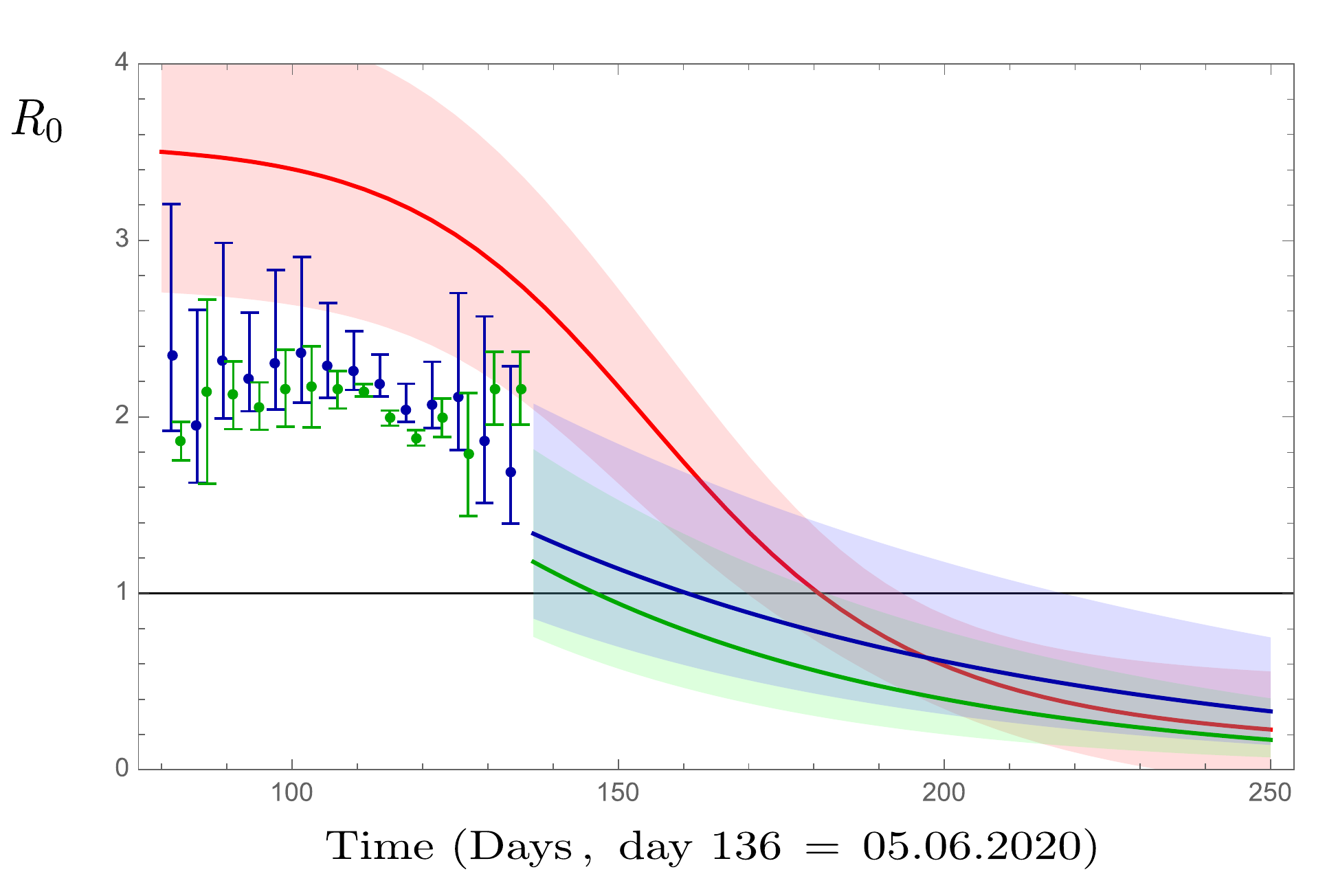}
\caption{\emph{Confirmed cases, deaths, active infections and $R_0$ versus time for Colombia
with data until $05.06.2020$. The red curves and error bars correspond to the prediction of the time independent SIR model. The green curves correspond to the prediction of a SIR model with time
dependent $\beta$ estimated from $(\gamma+I'/I)$ first by steps, and form an exponential fit. The blue curves correspond to the prediction of a SIR model with time
dependent $\beta$ estimated from $(\gamma+(I+R)'/(I+R))$ as an exponential fit.  The error bars give a in interval with $95\%$ confidence
interval.}}
\label{CO}
\end{center}
\end{figure}

\begin{figure}[h]
\begin{center}
\includegraphics[width=.48\textwidth]{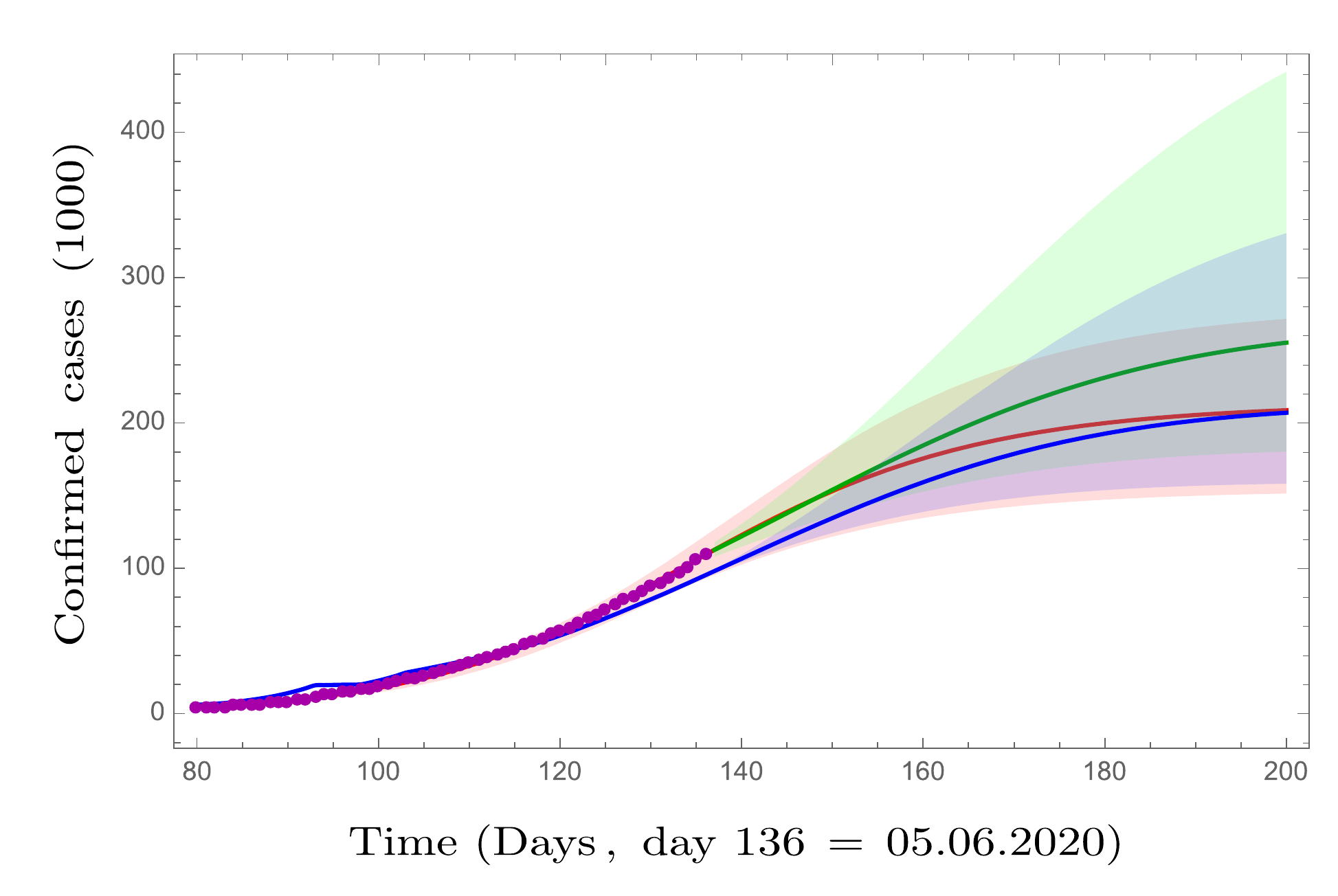}   
\includegraphics[width=.48\textwidth]{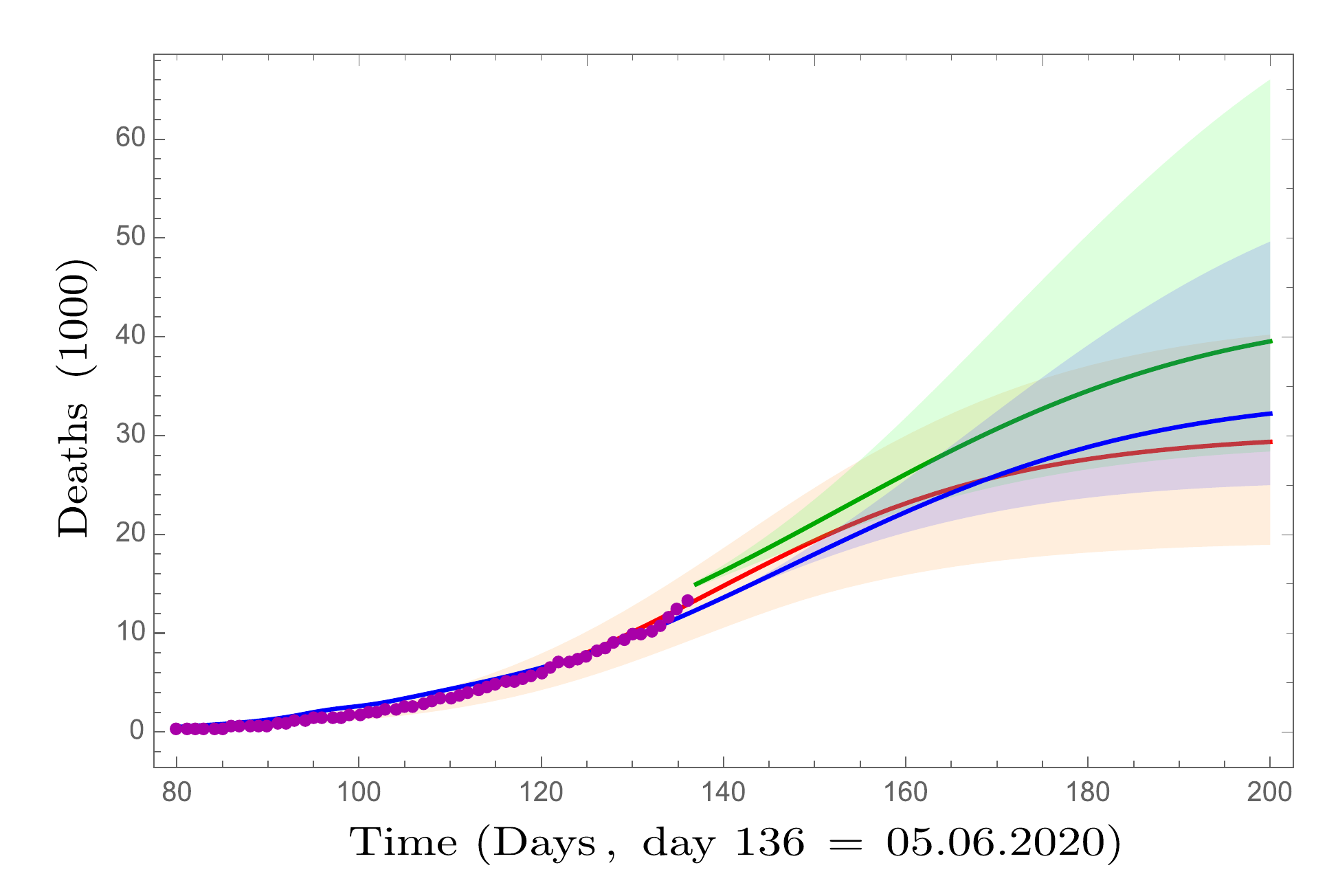}   
\includegraphics[width=.48\textwidth]{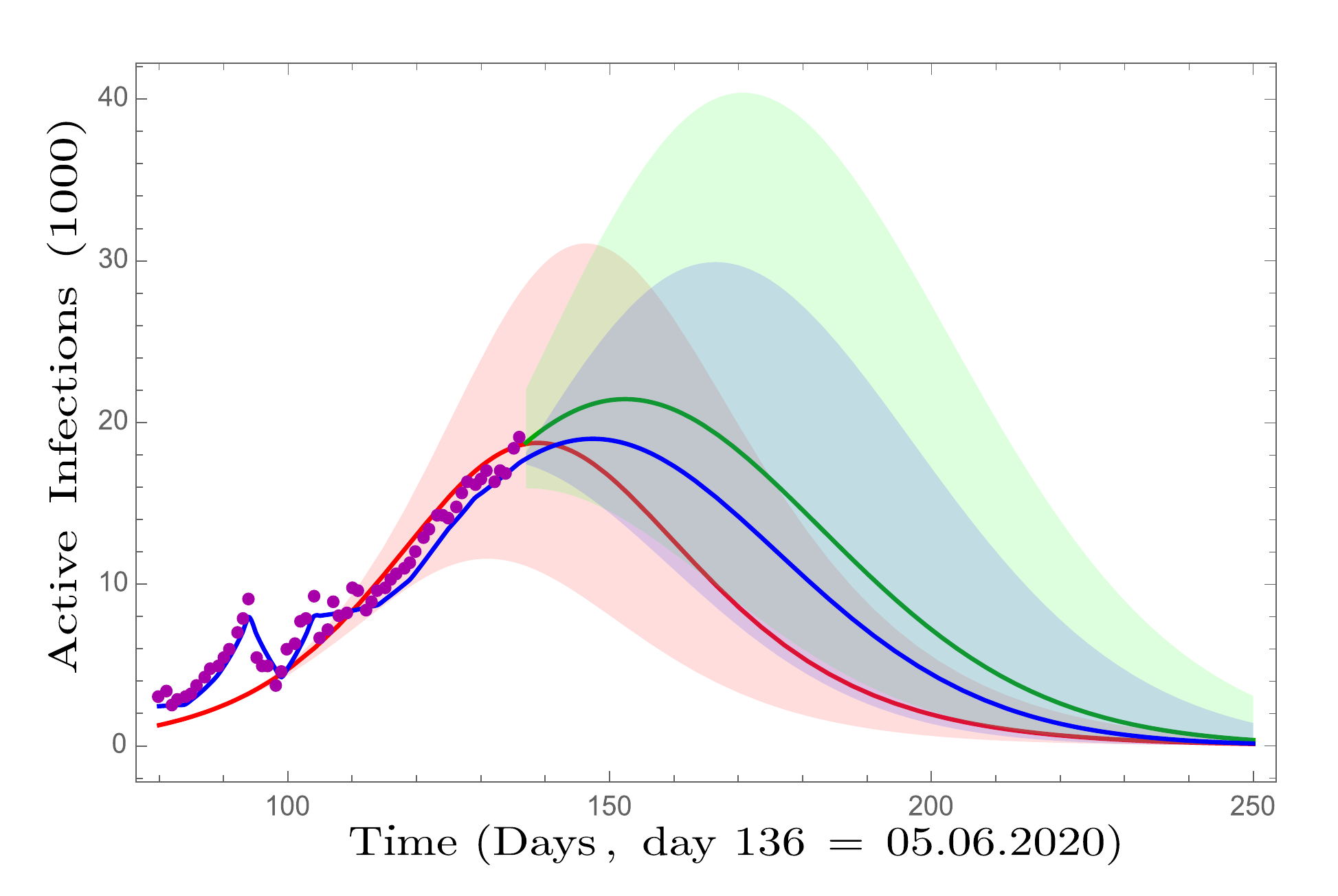}
\includegraphics[width=.48\textwidth]{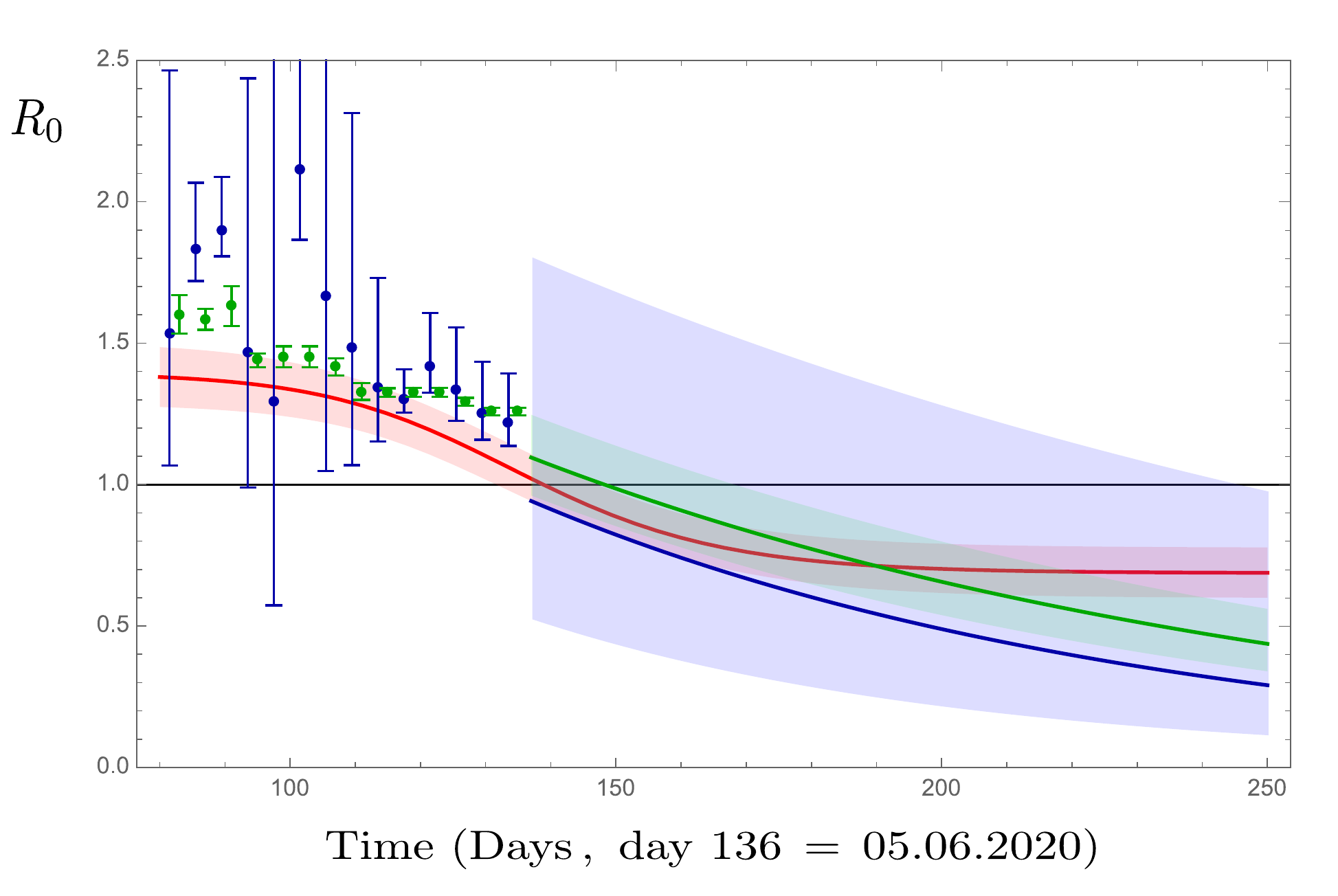}
\caption{\emph{Confirmed cases, deaths, active infections and $R_0$ versus time for Mexico
with data until $05.06.2020$. The red curves and error bars correspond to the prediction of the time independent SIR model. The green curves correspond to the prediction of a SIR model with time dependent $\beta$ estimated from $(\gamma+I'/I)$ first by steps, and form an exponential fit. 
The blue curves correspond to the prediction of a SIR model with time
dependent $\beta$ estimated from $(\gamma+(I+R)'/(I+R))$ as an exponential fit. The error bars give a in interval with $95\%$ confidence
interval.}}
\label{MX}
\end{center}
\end{figure}

\begin{figure}[h]
\begin{center}
\includegraphics[width=.48\textwidth]{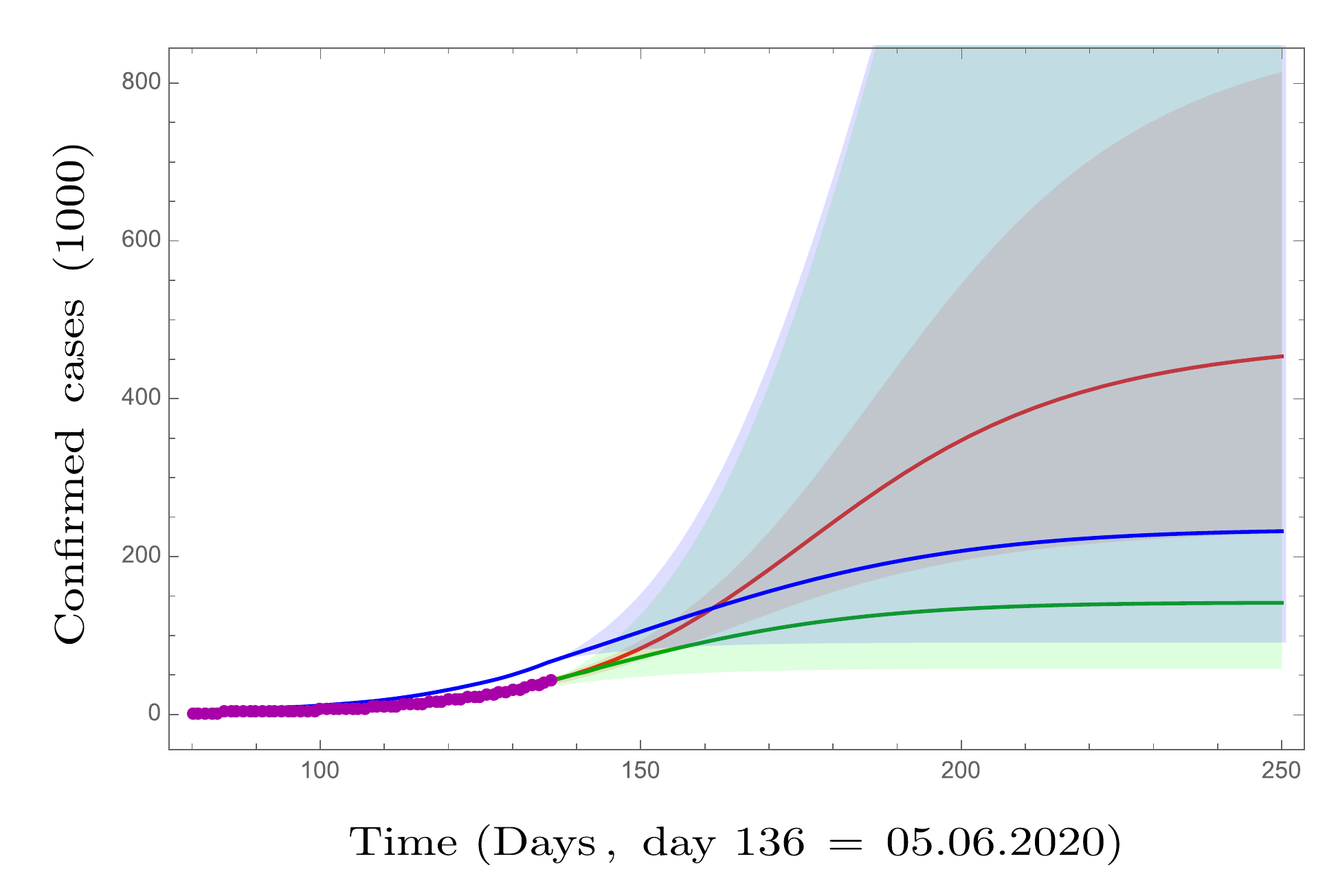}   
\includegraphics[width=.48\textwidth]{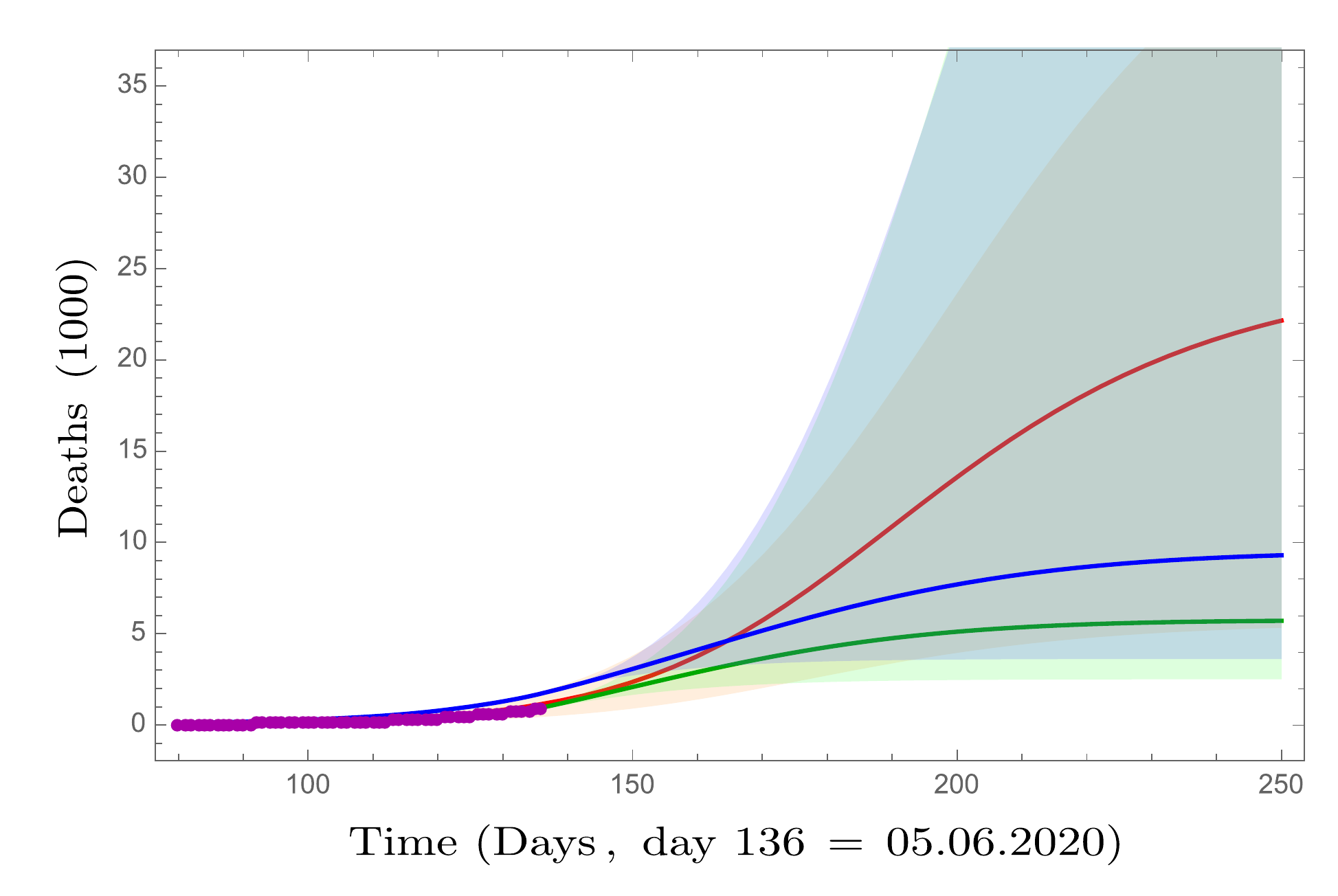}   
\includegraphics[width=.48\textwidth]{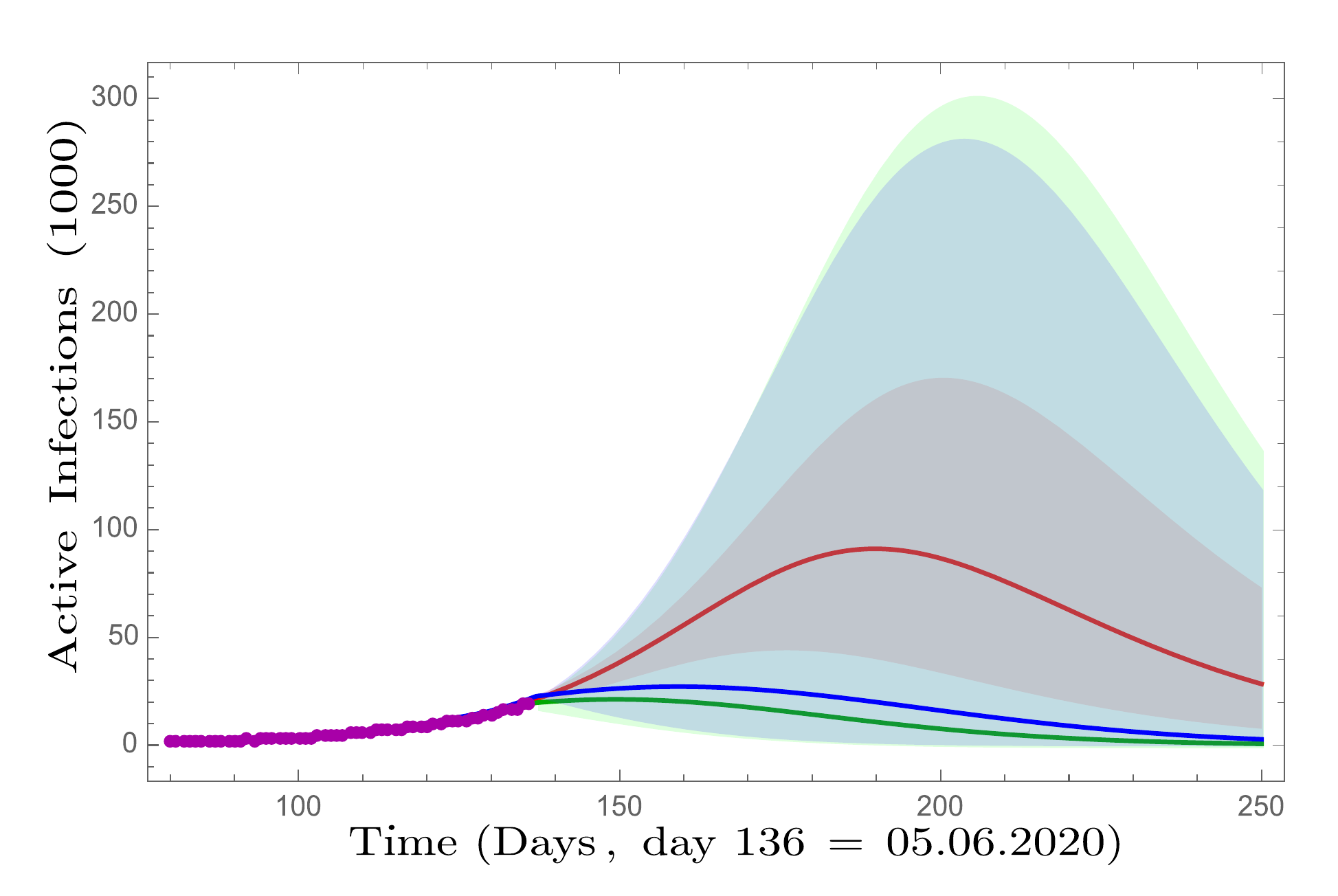}
\includegraphics[width=.48\textwidth]{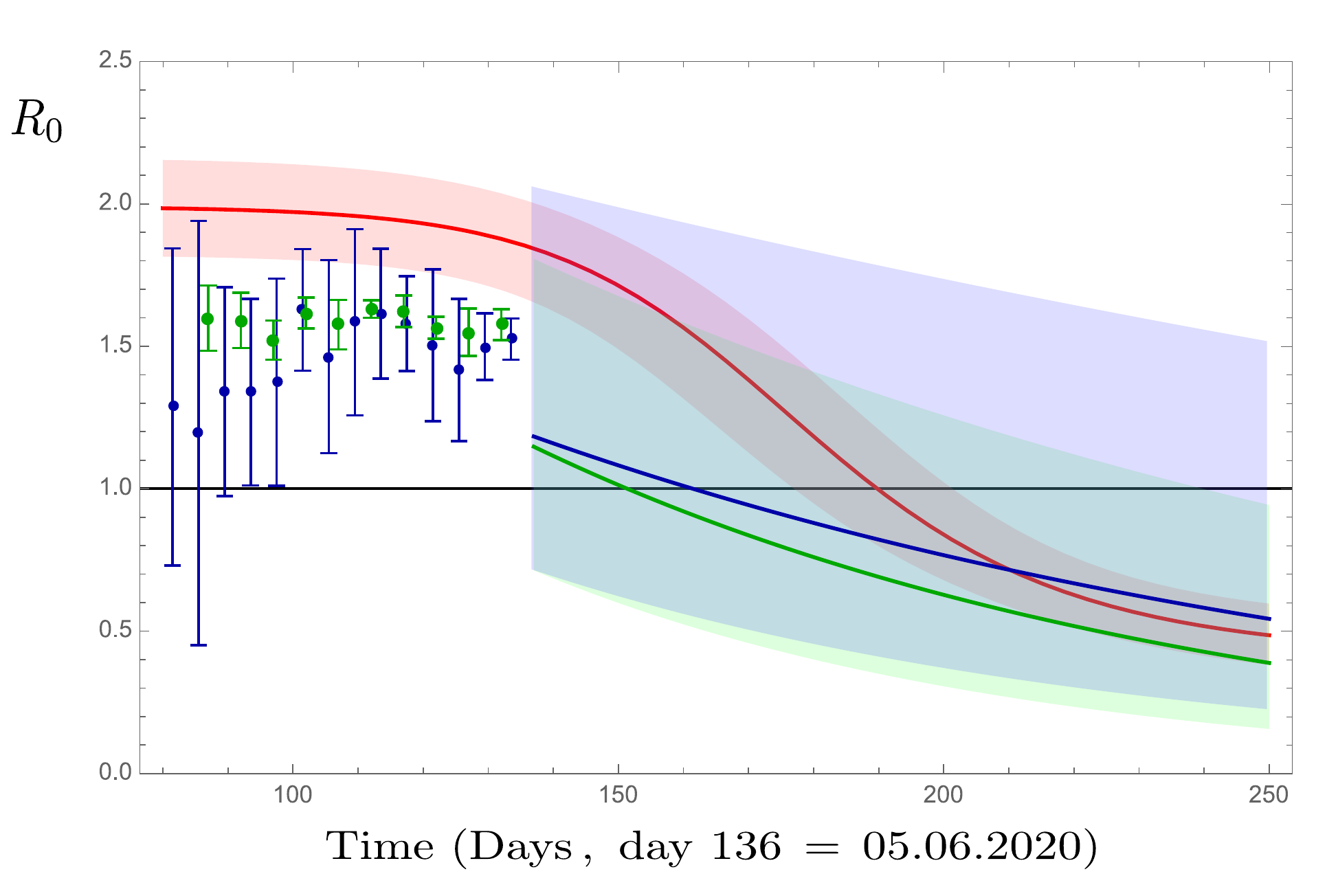}
\caption{\emph{Confirmed cases, deaths, active infections and $R_0$ versus time for South Africa
with data until $05.06.2020$. The red curves and error bars correspond to the prediction of the time independent SIR model. The green curves correspond to the prediction of a SIR model with time
dependent $\beta$ estimated from $(\gamma+I'/I)$ first by steps, and form an exponential fit. The blue curves correspond to the prediction of a SIR model with time
dependent $\beta$ estimated from $(\gamma+(I+R)'/(I+R))$ as an exponential fit. The error bars give a in interval with $95\%$ confidence
interval.}}
\label{ZA}
\end{center}
\end{figure}

Assuming the linear regime (\ref{eq:I3}) we take into consideration that the
contagion rate varies in time. To describe the epidemia in previous times
to the present we assume a $\beta$ which is changing
by pieces in intervals of 4-6 days. Locally in time we make a fit of the $I(t)$ versus $t$
data employing the linear regime solution (\ref{linEQ}).
This is  locally we consider  intervals of constant $\beta$. In  the Tables  \ref{TbAR}, \ref{TbBR}, \ref{TbCO}, 
\ref{TbMX} and \ref{TbZA}  we summarize the values of $\beta, R_0$ and their errors for a $95\%$ percent
interval of confidence for the studied countries, also till the date 05.06.20.  In the Figures
\ref{AR}, \ref{BR}, \ref{CO}, \ref{MX}, \ref{ZA} we can appreciate the changes in $R_0$ for this model 
changing by pieces. Also one can observe the curves of the populations evolution
for the different countries. 

We estimate the local values of $\beta$ from two different formulas, which are valid
only in the linear regime. The first of them uses the data of the detected accumulated cases:
\begin{eqnarray}
\beta(t)=\frac{(I(t)+R(t))'}{(I(t)+R(t))}+\gamma.\label{beta1}
\end{eqnarray}
The second uses the data for the detected active cases:
\begin{eqnarray}
\beta(t)=\frac{I(t)'}{I(t)}+\gamma.\label{beta2}
\end{eqnarray}
This last one has been used in \cite{26}. As mentioned they are equivalent to  $R_0^{(1)}$ in (\ref{R01}) and $R_0^{(2)}$ in (\ref{R02}). Figures \ref{AR}, \ref{BR} \ref{CO}, \ref{MX}, \ref{ZA} show with blue dots and error bars the estimates for (\ref{beta2}),
and with a blue line the populations obtained from evolving the SIR with (\ref{beta2}), this is done till the last data point.
With green dots and error bars they show only the estimate for (\ref{beta1}) till  $05.06.20$. The figures are organized as follows:
The sub-figures 1,2,3 and 4 show the dependences $\cI(t)+\cR(t), \cD(t), \cI(t)$ and $R_0(t)$.
The quantity $\cD(t)$ representing the evolution of the deaths is not considered in the equations of the SIR model of this section.
However we estimate it considering it as a proportion of the accumulated cases $\mathcal{C}(t)$, this is $\cD(t)=\mu  \mathcal{C}(t)$.
To estimate it we do a linear fit of $\cD(t)$ versus $ \mathcal{C}(t)$ taking all
the data points from the $16.03.20$ to the $05.06.20$.

The last sub-figure of the mentioned figures denotes the step determined values
of $R_0=\beta(t)/\gamma$ with the error bars, for the dates till $05.06.20$. After
$05.06.20$ we extrapolate with an exponential dependence.
Both equations (\ref{beta1}) and (\ref{beta2}) are approximated but exact in the linear regime $N\gg 1$, however in the measurements
(\ref{beta1}) has less uncertainty than (\ref{beta2}). This can be seen comparing
the error bars in the plots of $R_0$: the green bars  for (\ref{beta1}) and the blue bars for (\ref{beta2}). The intention of these
two approaches is to obtain independent but comparable estimates of the time variation of
the contagion rate.

\subsection{Contagion rate with exponential decay in time}
\label{s33}

In this subsection we consider the future predictions of the models with variable contagion rate $\beta(t)$,
taking into account an exponential time dependence. This dependence was proposed and explored for COVID-19
outbreak in \cite{26}. In that work the curve for the active cases of Cuba was correctly described,
and the estimated peak differed only by few days with the actual one. It has as well
been explored successfully to describe the evolution of the pandemia in Italy \cite{27}. In the linear regime
of the differential equations the maximum of the active cases is achieved
when the condition $R_0=1$ is reached. All the studied countries
are still in the region $R_0>1$. Here we apply the mentioned dependence to the studied countries.
However in the evolution for some cases the exponential fit for $\beta(t)$ is more appropriated, for others even-so a fit
to the exponential can be performed in a long period of time
$\beta$ values have not been reduced considerably.  Therefore 
these forecasts have to be taken with care, as they assume
the country will have in the future similar confinement measures as
in the interval fitted.

The dependence for $\beta(t)$ that we will fit to the data reads:
\begin{eqnarray}
\beta(t)= \beta_0 \text{exp}(-b_0 t). \label{expB}
\end{eqnarray}
We have estimated this dependence in two different ways. Considering the regime of $N$ large, which in our case is the population
from the different countries we have made a fit of the curves 
$R_0^{(1)}=\frac{(I+R)'}{\gamma(I+R)}+1$ vs. t (\ref{beta1}) and $R_0^{(2)}=\frac{I'}{\gamma I}+1$ vs. $t$ (\ref{beta2}). These approaches are accurate
in the linear regime, with differential equations shown in (\ref{linEQ}).
They render different estimates of $\beta(t)$ shown in the figures
 \ref{AR}, \ref{BR}, \ref{CO}, \ref{MX} and \ref{ZA},  and both are in agreement.

The $R_0(t)=\beta(t)/\gamma$ values obtained from  adjusting $\frac{(I+R)'}{(I+R)}+\gamma$ vs. t.
are represented in the last plots of the figures. The data points considered
for the fits are the dates from $16.03.20$ to $5.06.20$. The exponential
dependence is extrapolated from $6.06.2020$ into the future. The blue lines and blue colored region
represent the media and confidence intervals for the $R_0$ obtained with (\ref{beta2}); while the green lines and the green colored regions
represent the media and the confidence intervals for the $R_0$ obtained with (\ref{beta1}). The sub-figures from the 1st to the 3rd
give the populations $(\cI+\cR)(t)$, $\cD(t)$ and $\cI(t)$, the notation for the exponential
contagion rate forecast is the same as in the plot $R_0(t)$ i.e. the 
lines are the SIR evolution with the mean values of $\beta(t)$ taken from  ((\ref{beta1}), green line)
and  ((\ref{beta2}), blue line), and the colored regions represent the confidence intervals
of the populations   ((\ref{beta1}), green region) and ((\ref{beta1}), blue region).
In table  \ref{tab:PN} we summarize the different values of the exponential $\beta(t)$ with confidence intervals 
for the different countries.

In the plot for $R_0$ the moment of the active cases $\cI(t)$ peak is reached when the mean curve crosses the read line representing 1,
this is $R_0=1$. This is the moment when $\beta(t)$ reaches the mean value of $\gamma$.  In Table \ref{picos}
we show the dates predicted for the peaks, the number of active cases in the peak,  the number of total infected 
and the estimates for the total deaths of the studied countries for an exponential $\beta(t)$, they are shown
with uncertainties. As discussed the estimates are obtained from two approaches.

\section{Analysis and Conclusions}
\label{s4}

In this work we have explored the SIR model following two different approaches. In the first case we opted for a fixed contagion rate. The effects of contention measures are reflected on an effective value of the total 
susceptible population $N_{eff}$ (much smaller than the actual country population) and incorporates an overall time delay in the evolution of the differential equations.

In the second approach the population $N$ was taken to be the population of the country in consideration and the contention measures are reflected on a time varying contagion rate $\beta(t)$. This contagion rate is considered an exponential decay motivated by the previous study for COVID-19 \cite{26}, and has been also considered in \cite{27}, and it reflects the effect
of the contention measures. Both of these approaches are fully self contained, meaning that we estimate all the parameters involved.
The evolutions described are the ones of the detected populations $\cI(t)$ and $\cR(t)$. If one would like to estimate the real infections is necessary to rely on the values of the detection rates $\alpha$ 
reported in the literature \cite{5,12}.

Both approaches can be regarded as complementary ways to tackle the same problem. In  each situation and despite the specific reporting and testing conditions of each country, we observe that both approaches lead to similar results and agree within error bars. This also highlights the versatility of the SIR model and its effectiveness in modeling this type of situations. 
The agreement of the approaches comes from the intuition that the reduced mobility of the society could be casted into the SIR model in two different but complementary ways: either by effectively reducing in the pool of interaction ($N_{eff}$) or by a time decreasing contagion rate ($\beta(t)$). In section \ref{sec:sir} we have developed the mathematical machinery to compare between the parameters involved in each of the approaches, we also observe a good agreement among these quantities as can be indeed seen in Table \ref{picos}. \\

The results of this work show the validity of the SIR model and its modified versions in order
to describe the spread of communicable diseases with reasonable accuracy, in particular the novel coronavirus outbreak. 
The versatility and 
simplicity of the model 
permits us to develop a ``personalized" analysis of the different countries situations. For some cases, we have found universality of the parameters, as in the case of the recovery rate $\gamma$ whose values coincide for most of the countries under study.
We would like to emphasize that the time dependent models considered 
allow us to explore the time evolution of the pandemic, and to implement the effect of the restrictive mobility in the set of differential
equations. Similarly, the time independent model also proved to be useful in this description. It would be interesting to contrast the time dependent parameters with the corresponding NPIs put in place by each country over various periods of time. Due to the comparable results obtained though these approaches it would be interesting to explore this ``complementarity" of SIR models with a time decaying contagion rate $\beta$ and a large country population $N$ and the SIR models with an effective population $N_{eff}$. We hope to come back to these issues in a future work. \\

The results and observations made in this work have to be taken carefully due to many reasons, among which one has to underline the fact that many countries are ending their lockdown measures before the peak of active infections is reached. This can significantly increase the number of infections and in some cases they might exceed the 95\% confidence levels established in this work. Also from estimates concerning the total number of infections as well as the effective pool of susceptible $N_{eff}$, we observe that at the end of the peak no heard immunity would be reached in any of the cases studied. This is the way in which the epidemia has been developed in all the countries that have control it so far.  Also in our results we observe a certain variability on the peak of active infections. 
Additionally a second peak of the outbreak could arise, the eventuality of this new wave of spreading has not been contemplated in this work. Nevertheless we consider that our analysis could contribute to those explorations.



\begin{table}[h]
\centering
\renewcommand{\arraystretch}{1.4}
{\small
\begin{tabular}{|c|c|c|c|c|}\hline
INTERVAL & $R^{(1)}_0$ & $\Delta R^{(1)}_{0}$& $R^{(2)}_0$ & $\Delta R^{(2)}_{0}$\\ \hline\hline
\text{03.16.20 - 03.20.20} & 6.39 & 0.83 & 8.32 & 1.66 \\
 \text{03.20.20 - 03.25.20} & 4.26 & 2.03 & 7.93 & 3.61 \\
 \text{03.25.20 - 03.30.20} & 2.05 & 1.55 & 5.63 & 1.38 \\
 \text{03.30.20 - 04.04.20} & 3.23 & 0.91 & 4.41 & 1.23 \\
 \text{04.04.20 - 04.09.20} & 1.98 & 0.19 & 2.61 & 0.35 \\
 \text{04.09.20 - 04.14.20} & 1.27 & 0.41 & 2.56 & 0.50 \\
 \text{04.14.20 - 04.19.20} & 2.03 & 0.19 & 2.46 & 0.35 \\
 \text{04.19.20 - 04.24.20} & 1.64 & 0.40 & 2.72 & 0.46 \\
 \text{04.24.20 - 04.29.20} & 1.81 & 0.11 & 2.14 & 0.12 \\
 \text{04.29.20 - 05.04.20} & 1.53 & 0.11 & 1.91 & 0.1 \\
 \text{05.04.20 - 05.09.20} & 2.02 & 0.06 & 2.19 & 0.1 \\
 \text{05.09.20 - 05.14.20} & 1.46 & 0.31 & 2.48 & 0.06 \\
 \text{05.14.20 - 05.19.20} & 2.43 & 0.1 & 2.41 & 0.12 \\
 \text{05.19.20 - 05.24.20} & 3.05 & 0.22 & 3.23 & 0.10 \\
 \text{05.24.20 - 05.29.20} & 2.58 & 0.09 & 2.72 & 0.07 \\
 \text{05.29.20 - 06.05.20} & 2.42 & 0.13 & 2.54 & 0.07 \\ \hline
\end{tabular}
}
 \caption{ \label{TbAR}   Estimates for the variable contagion rate for Argentina.
Values obtained from $(\frac{(I+R)'}{\gamma(I+R)}+1)$ and
$(\frac{I'}{\gamma I}+1)$ .   Still the condition $R_{0}=\beta/\gamma<1$,is not reached.\label{R0AR}}
\end{table}

\begin{table}[h]
\centering
\renewcommand{\arraystretch}{1.4}
{\small
\begin{tabular}{|c|c|c|c|c|}\hline
INTERVAL & $R^{(1)}_0$ & $\Delta R^{(1)}_{0}$& $R^{(2)}_0$ & $\Delta R^{(2)}_{0}$\\ \hline\hline
\text{03.16.20 - 03.21.20} & 5.13 & 0.42 & 5.73 & 0.79 \\
 \text{03.21.20 - 03.27.20} & 2.90 & 0.08 & 3.04 & 0.19 \\
 \text{03.27.20 - 04.02.20} & 3.08 & 0.16 & 3.37 & 0.26 \\
 \text{04.02.20 - 04.08.20} & 2.55 & 0.1 & 2.78 & 0.19 \\
 \text{04.08.20 - 04.14.20} & 1.29 & 0.70 & 2.18 & 0.16 \\
 \text{04.14.20 - 04.20.20} & 0.15 & 0.51 & 1.99 & 0.14 \\
 \text{04.20.20 - 04.26.20} & 2.69 & 0.07 & 2.19 & 0.08 \\
 \text{04.26.20 - 05.02.20} & 1.78 & 0.15 & 1.97 & 0.13 \\
 \text{05.02.20 - 05.08.20} & 2.05 & 0.08 & 2.07 & 0.12 \\
 \text{05.08.20 - 05.14.20} & 1.83 & 0.12 & 1.98 & 0.1 \\
 \text{05.14.20 - 05.20.20} & 1.74 & 0.05 & 1.95 & 0.08 \\
 \text{05.20.20 - 05.26.20} & 1.67 & 0.04 & 1.71 & 0.07 \\
 \text{05.26.20 - 06.02.20} & 1.62 & 0.05 & 1.71 & 0.08 \\ \hline
\end{tabular}
}
 \caption{ \label{TbBR}   Estimates for the variable contagion rate for Brazil.
Values obtained from $(\frac{(I+R)'}{\gamma(I+R)}+1)$ and
$(\frac{I'}{\gamma I}+1)$ .   Still the condition $R_{0}=\beta/\gamma<1$,is not reached.\label{R0BR}}
\end{table}

\begin{table}[h]
\centering
\renewcommand{\arraystretch}{1.4}
{\small
\begin{tabular}{|c|c|c|c|c|}\hline
INTERVAL & $R^{(1)}_0$ & $\Delta R^{(1)}_{0}$& $R^{(2)}_0$ & $\Delta R^{(2)}_{0}$\\ \hline\hline
\text{03.16.20 - 03.19.20} & 2.63 & 1.65 & 5.75 & 3.79 \\
 \text{03.19.20 - 03.23.20} & 4.72 & 0.947 & 6.26 & 1.91 \\
 \text{03.23.20 - 03.27.20} & 2.25 & 1.11 & 4.11 & 2.07 \\
 \text{03.27.20 - 03.31.20} & 3.54 & 0.113 & 3.88 & 0.121 \\
 \text{03.31.20 - 04.04.20} & 2.53 & 0.309 & 3.27 & 0.485 \\
 \text{04.04.20 - 04.08.20} & 2.24 & 0.407 & 3.17 & 0.830 \\
 \text{04.08.20 - 04.12.20} & 1.81 & 0.345 & 2.72 & 0.622 \\
 \text{04.12.20 - 04.16.20} & 1.18 & 0.0835 & 1.86 & 0.109 \\
 \text{04.16.20 - 04.20.20} & 1.42 & 0.243 & 2.14 & 0.521 \\
 \text{04.20.20 - 04.24.20} & 1.90 & 0.0883 & 2.12 & 0.191 \\
 \text{04.24.20 - 04.28.20} & 1.80 & 0.0975 & 2.06 & 0.134 \\
 \text{04.28.20 - 05.02.20} & 1.75 & 0.166 & 2.16 & 0.219 \\
 \text{05.02.20 - 05.06.20} & 1.86 & 0.108 & 2.17 & 0.230 \\
 \text{05.06.20 - 05.10.20} & 2.00 & 0.0712 & 2.15 & 0.106 \\
 \text{05.10.20 - 05.14.20} & 2.09 & 0.0393 & 2.15 & 0.0349 \\
 \text{05.14.20 - 05.18.20} & 1.98 & 0.0399 & 1.99 & 0.0431 \\
 \text{05.18.20 - 05.22.20} & 1.82 & 0.0316 & 1.88 & 0.0436 \\
 \text{05.22.20 - 05.26.20} & 1.81 & 0.0930 & 1.99 & 0.109 \\
 \text{05.26.20 - 05.30.20} & 1.22 & 0.204 & 1.79 & 0.348 \\
 \text{05.30.20 - 06.5.20} & 1.10 & 0.149 & 2.16 & 0.207 \\ \hline
\end{tabular}
}
 \caption{\label{TbCO}  Estimates for the variable contagion rate for Colombia.
Values obtained from $(\frac{(I+R)'}{\gamma(I+R)}+1)$ and
$(\frac{I'}{\gamma I}+1)$ .   Still the condition $R_{0}=\beta/\gamma<1$,is not reached.\label{R0CO}}
\end{table}

\begin{table}[h]
\centering
\renewcommand{\arraystretch}{1.4}
{\small
\begin{tabular}{|c|c|c|c|c|}\hline
INTERVAL & $R^{(1)}_0$ & $\Delta R^{(1)}_{0}$& $R^{(2)}_0$ & $\Delta R^{(2)}_{0}$\\ \hline\hline
 \text{03.16.20 - 03.20.20} & 2.26 & 0.20 & 2.69 & 0.36 \\
 \text{03.20.20 - 03.25.20} & 1.92 & 0.11 & 2.08 & 0.21 \\
 \text{03.25.20 - 03.30.20} & 1.88 & 0.12 & 2.09 & 0.20 \\
 \text{03.30.20 - 04.04.20} & 0.37 & 0.34 & 1.74 & 0.03 \\
 \text{04.04.20 - 04.09.20} & 1.98 & 0.07 & 1.83 & 0.1 \\
 \text{04.09.20 - 04.14.20} & 0.60 & 0.23 & 1.60 & 0.07 \\
 \text{04.14.20 - 04.19.20} & 1.60 & 0.06 & 1.58 & 0.04 \\
 \text{04.19.20 - 04.24.20} & 1.75 & 0.04 & 1.63 & 0.07 \\
 \text{04.24.20 - 04.29.20} & 0.74 & 0.44 & 1.44 & 0.02 \\
 \text{04.29.20 - 05.04.20} & 1.61 & 0.13 & 1.45 & 0.04 \\
 \text{05.04.20 - 05.09.20} & 0.43 & 0.31 & 1.42 & 0.03 \\
 \text{05.09.20 - 05.14.20} & 0.73 & 0.17 & 1.33 & 0.03 \\
 \text{05.14.20 - 05.19.20} & 1.19 & 0.02 & 1.33 & 0.02 \\
 \text{05.19.20 - 05.24.20} & 1.23 & 0.05 & 1.33 & 0.02 \\
 \text{05.24.20 - 05.29.20} & 1.12 & 0.06 & 1.29 & 0.01 \\
 \text{05.29.20 - 06.5.20} & 1.05 & 0.04 & 1.26 & 0.01 \\ \hline
\end{tabular}
}
 \caption{ \label{TbMX}  Estimates for the variable contagion rate for Mexico.
Values obtained from $(\frac{(I+R)'}{\gamma(I+R)}+1)$ and
$(\frac{I'}{\gamma I}+1)$ .   Still the condition $R_{0}=\beta/\gamma<1$,is not reached.\label{R0MX}}
\end{table}
 
\begin{table}[h]
\centering
\renewcommand{\arraystretch}{1.4}
{\small
\begin{tabular}{|c|c|c|c|c|}\hline
INTERVAL & $R^{(1)}_0$ & $\Delta R^{(1)}_{0}$& $R^{(2)}_0$ & $\Delta R^{(2)}_{0}$\\ \hline\hline
\text{03.16.20 - 03.20.20} & 2.81 & 0.77 & 4.35 & 1.12 \\
 \text{03.20.20 - 03.25.20} & 3.12 & 0.28 & 3.88 & 0.47 \\
 \text{03.25.20 - 03.30.20} & 1.48 & 0.36 & 2.06 & 0.67 \\
 \text{03.30.20 - 04.04.20} & 1.16 & 0.06 & 1.38 & 0.09 \\
 \text{04.04.20 - 04.09.20} & 1.34 & 0.04 & 1.41 & 0.08 \\
 \text{04.09.20 - 04.14.20} & 0.73 & 0.28 & 1.47 & 0.12 \\
 \text{04.14.20 - 04.19.20} & 0.45 & 0.37 & 1.59 & 0.12 \\
 \text{04.19.20 - 04.24.20} & 1.15 & 0.12 & 1.62 & 0.10 \\
 \text{04.24.20 - 04.29.20} & 0.87 & 0.21 & 1.49 & 0.08 \\
 \text{04.29.20 - 05.04.20} & 1.41 & 0.11 & 1.63 & 0.04 \\
 \text{05.04.20 - 05.09.20} & 1.12 & 0.17 & 1.56 & 0.10 \\
 \text{05.09.20 - 05.14.20} & 1.30 & 0.16 & 1.63 & 0.02 \\
 \text{05.14.20 - 05.19.20} & 1.47 & 0.07 & 1.64 & 0.06 \\
 \text{05.19.20 - 05.24.20} & 1.24 & 0.13 & 1.57 & 0.03 \\
 \text{05.24.20 - 05.29.20} & 1.17 & 0.12 & 1.54 & 0.09 \\
 \text{05.29.20 - 06.5.20} & 1.45 & 0.04 & 1.58 & 0.05 \\\hline
\end{tabular}
}
 \caption{\label{TbZA} Estimates for the variable contagion rate for South Africa.
Values obtained from $(\frac{(I+R)'}{\gamma(I+R)}+1)$ and
$(\frac{I'}{\gamma I}+1)$ .   Still the condition $R_{0}=\beta/\gamma<1$,is not reached.\label{R0ZA}}
\end{table}

\newpage 

\section*{Acknowledgements}

This work is dedicated to the loving memory of Mar\'{\i}a Eva Lozada de Pe\~{n}a. We thank Juan Barranco, Argelia Bernal,  Milagros Bizet, Alejandro Cabo Bizet, Alejandro Cabo Montes de Oca, Magda Lorena Forero Pe\~{n}a, Alma Gonz\'alez,  Oscar Loaiza, Albrecht Klemm, Mauro Napsuciale, Gustavo Niz, Octavio Obreg\'on, Miguel Sabido, Matthias Schmitz and  Luis Ure\~{n}a, for useful discussions and suggestions. We also thank our family and friends  for motivating us to carry this 
analysis for our countries. We
thank the project CONACyT A1-S- 37752, UG Project CIIC 290/2020, Project COVID19-UG 36/2020 “Modelaci\'on matem\'atica de la propagaci\'on del COVID-19 en M\'exico y Guanajuato” and the Data Lab of the University of Guanajuato.  
DKMP is supported by the Simons Foundation Mathematical and Physical Sciences Targeted Grants to Institutes, Award ID:509116.





\end{document}